\documentclass[useAMS,usenatbib]{mn2e}

\usepackage{graphicx}
\usepackage{amssymb}

\usepackage{natbib,times}
\usepackage{lscape}

\newcommand{\mc}{\multicolumn}

\begin{document}

\title[Dynamical state of 2092 rich clusters]
{Substructure and dynamical state of 2092 rich clusters of galaxies 
derived from photometric data}

\author[Z. L. Wen and J. L. Han]
{Z. L. Wen\thanks{E-mail: zhonglue@nao.cas.cn} 
and J. L. Han 
\\
National Astronomical Observatories, Chinese Academy of Sciences, 20A
Datun Road, Chaoyang District, Beijing 100012, China\\
}

\date{Accepted 2013 ... Received 2013 ...}

\pagerange{\pageref{firstpage}--\pageref{lastpage}} \pubyear{2013}

\maketitle

\label{firstpage}


\begin{abstract}
Dynamical state of galaxy clusters is closely related to their
observational properties in X-ray, optical and radio wavelengths. We
develop a method to diagnose the substructure and dynamical state of
galaxy clusters by using photometric data of Sloan Digital Sky Survey
(SDSS). To trace mass distribution, the brightness distribution of
member galaxies is smoothed by using a Gaussian kernel with a weight
of their optical luminosities. After deriving the asymmetry, the ridge
flatness and the normalized deviation of the smoothed optical map, we
define a relaxation parameter, $\Gamma$, to quantify dynamical state
of clusters. This method is applied to a test sample of 98 clusters of
$0.05<z\lesssim0.42$ collected from literature with known dynamical
states and can recognize dynamical state for relaxed ($\Gamma\ge0$)
and unrelaxed ($\Gamma<0$) clusters with a success rate of 94\%.
We then calculate relaxation parameters of 2092 rich clusters
previously identified from the SDSS, of which 28\% clusters are
dynamically relaxed with $\Gamma\ge0$. We find that the dominance and
absolute magnitude of the brightest cluster galaxies closely correlate
with dynamical state of clusters. The emission power of radio halos
is quantitatively related to cluster dynamical state, beside the known
dependence on the X-ray luminosity.
\end{abstract}


\section{Introduction}

Clusters of galaxies are the most massive bound systems in the
universe. The hierarchical model \citep{pee80} predicts that clusters
of galaxies form by accretion and merging of smaller sub-clusters and
groups \citep{cwj+99}. Dynamical state of clusters can usually be
divided into two broad classes, relaxed and unrelaxed. Determination
of cluster dynamical state is important not only for understanding
cluster properties in X-ray, optical and radio wavelengths
\citep{srt+08,fgg+12,me12} but also for cosmological studies
\citep{see+03,pb07,cmn10}.

Dynamical state of a few hundred clusters has been diagnosed based on
substructures in X-ray image and spectra data
\citep[e.g.,][]{sks+05,vmm+05,ars+08,me12}, quantitatively by using
the power ratio \citep[e.g.,][]{bt95,bpa+10}, the centroid shift
\citep[e.g.,][]{mef+95,mjf+08,me12}, the asymmetry and the
concentration \citep[e.g.,][]{hbh+07,srt+08}. About 40\%--70\% of
clusters show various substructures in X-ray images
\citep[e.g.,][]{jf99,sbr+01,sks+05} which are unrelaxed clusters
undergone multiple merger events in the recent past
\citep{me12}. Relaxed clusters generally have cool cores of
intracluster gas in the center
\citep{esf92,afj+01,bfs+05,vmm+05,ars+08}, and unrelaxed clusters are
systematically hotter than relaxed clusters \citep{om02,sks+05,pd07}.
Because the hydrostatic equilibrium is violated in unrelaxed clusters
\citep{pb07}, clusters of various dynamical states have different
scaling relations, e.g., between cluster mass and X-ray luminosity or
temperature \citep{crb+07,all12}.

Dynamical state is related to the cluster properties in radio
\citep{bro+94,brb+98,lgg+10}. Almost all radio halos or radio relics
were detected from unrelaxed merging clusters and mini-halo were
usually found from relaxed clusters
\citep{sbr+01,gmv+04,ceg+10,fgg+12}.

In optical, three-dimensional distribution and motions of the member
galaxies are the most direct tracer of dynamical state of
clusters. Spectroscopic surveys of member galaxies are powerful to
reveal substructures along the line of sight \citep{ds88}. Relaxed
clusters should have a Gaussian distribution for redshifts of member
galaxies, and non-Gaussian redshift distribution is a clear evidence
of unrelaxed state \citep{cd96,hmp+04}. Previously substructures were
searched from the spectroscopic data by many methods, e.g., the
$\Delta$-statistics which measures the deviations of the local radial
velocity distribution from the global values \citep{ds88}, the
hierarchical clustering method \citep{sg96}, the skewness and kurtosis
of velocity distributions \citep{wb90,ssc99}, the Anderson--Darling,
Kolmogorov and $\chi^2$ tests \citep{hph+09} and multidimensional
normal mixture modelling \citep{ets+10}. About 30\%--70\% of clusters
have substructures shown in spectroscopic data
\citep{ds88,wb90,hph+09,as10,evn+12,hpw+12}.

However, spectroscopic observations usually are incomplete for cluster
member galaxies especially for faint galaxies and only available for a
very limited sample of galaxy clusters. Projected two-dimensional
distributions of member galaxies were also used to search for
substructures by, e.g., tests of the asymmetry, the angular separation
and density contrast \citep{wod88}, cluster centroid shift
\citep{kbp+01}, the adaptive-kernel based DEDICA algorithm
\citep{pis93,rbp+07}, the average two-point correlation function
statistic \citep{sgs93} and wavelet analysis \citep{fk06}. These
two-dimensional approaches work on positions of member galaxies and
have been applied to nearby clusters, which show that 30\%--70\% of
clusters have substructures \citep[e.g.,][]{fk06,rbp+07}.

Up to now, more than 130\,000 clusters have been identified from
optical data, e.g., from the Sloan Digital Sky Survey (SDSS)
\citep{whl09,hmk+10,spd+11,whl12} and thousands from X-ray data, e.g.,
from the {\it ROSAT} survey \citep{bvh+00,bsg+04}. Only a few hundred nearby
clusters have their substructures quantified from X-ray image or
optical spectrometry \citep[e.g.,][]{ds88,bt95,wbs+13}. The challenge
for quantifying dynamical state of a large sample of clusters is to
obtain deep X-ray observations or complete optical spectroscopic
redshifts of member galaxies.

In this paper, we develop a method to diagnose two-dimensional
substructures and dynamical states of galaxy clusters by using
multi-band optical photometric data, and quantify dynamical states for
a large sample of rich clusters. By using SDSS photometric redshifts,
we can identify luminous member galaxies of clusters with reasonable
completeness and diminish most contamination from foreground and
background galaxies \citep{whl09}.  Because more luminous member
galaxies trace more mass, the projected distribution of member
galaxies in the sky plane weighted with optical luminosity can well
trace the projected light and mass distribution of galaxy clusters
\citep{zbb+12,zbu+12}. Very relaxed clusters should have a very smooth
symmetrical mass and light distribution, while the unrelaxed clusters
should have substructures. To verify the method for quantification of
dynamical states from optical data, in Section 2 we collect a test
sample of rich clusters from literature with dynamical states
classified as relaxed and unrelaxed based on X-ray, optical and radio
data, and we obtain their optical data from SDSS. In Section 3, we
describe the method and define a relaxation parameter to quantify
cluster dynamical state based on the smoothed optical map for the
brightness distribution of member galaxies. We apply the method to the
test cluster sample, and find that the known relaxed and unrelaxed
clusters can be well distinguished with a success rate of 94\%. In
Section 4, we calculate the relaxation parameters for 2092 rich clusters
($0.05<z\le0.42$) taken from the catalog of \citet{whl12} which were
identified from optical photometric data of the SDSS-III. The
correlations between the relaxation parameter and cluster properties
are discussed. We present our conclusions in Section 5.

Throughout this paper, we assume a $\Lambda$CDM cosmology, taking
$H_0=$100 $h$ ${\rm km~s}^{-1}$ ${\rm Mpc}^{-1}$, with $h=0.72$, 
$\Omega_m=0.3$ and $\Omega_{\Lambda}=0.7$.

\begin{table*}
\caption[]{Test sample of clusters with known dynamical states in the field of SDSS-III}
\begin{center}
\resizebox{\textwidth}{!}{
\begin{tabular}{@{}lrrrccrrrl@{}}

\hline
\mc{1}{l}{Name}    &\mc{1}{c}{R.A.}       &\mc{1}{c}{Dec.}   &\mc{1}{c}{$z$}      & \mc{1}{c}{$r_{\rm BCG}$}& 
\mc{1}{c}{$r_{200}$}&\mc{1}{c}{$R_{L\ast}$}&\mc{1}{c}{$N_{200}$}&\mc{1}{c}{$\Gamma$} & \mc{1}{l}{Comment and reference}\\
                   &\mc{1}{c}{(deg)}      &\mc{1}{c}{(deg)}   &                    &                        &
\mc{1}{c}{(Mpc)}    &                     &                    &                    &                    \\
(1) & \mc{1}{c}{(2)} & \mc{1}{c}{(3)} & \mc{1}{c}{(4)} & (5) & (6) & \mc{1}{c}{(7)} & (8) & \mc{1}{c}{(9)}& (10) \\
\hline

RXC J0003.8+0203  &   0.95698 &    2.06647&  0.0978&  14.09&  1.57&  70.32&   46& $  0.28\pm0.04$&R, relaxed[wbs+13]\\
A13              &   3.41056 &$-19.50015$&  0.0943&  14.61&  1.78&  76.22&   46& $ -0.46\pm0.09$&U, merger[jsc+08],radio[fgg+12]\\   
CL0024.0+1652    &   6.64862 &   17.16197&  0.3871&  17.74&  1.87&  94.77&   72& $  0.49\pm0.04$&U, merger (line of sight)[cmk+02]\\   
A68              &   9.27851 &    9.15670&  0.2861&  16.72&  1.96& 107.22&   76& $ -0.50\pm0.10$&U, merger[st08]\\		    
A85              &  10.46029 &$ -9.30313$&  0.0554&  13.31&  1.68&  81.23&   67& $ -0.91\pm0.09$&U, merger,cool[ksr02],radio[fgg+12]\\
A2813            &  10.85232 &$-20.62496$&  0.2924&  17.15&  2.00& 133.15&   74& $ -0.24\pm0.04$&U, merger[me12]\\   
A98              &  11.60318 &   20.62172&  0.1028&  14.75&  1.75& 115.93&   98& $ -0.69\pm0.10$&U, merger[bro+94]\\	    
A115             &  14.00108 &   26.34230&  0.1920&  16.44&  2.05& 134.95&  104& $ -0.76\pm0.11$&U, merger[fbb+81],radio[fgg+12],cool[bfs+05]\\   
MACS J0140.0$-$0555&25.00334 &$ -5.91744$&  0.4474&  18.42&  1.76&  92.27&   68& $  0.00\pm0.08$&U, merger[me12]\\				    
A267             &  28.17485 &    1.00709&  0.2297&  15.59&  1.92& 130.44&  105& $  0.48\pm0.02$&U, merger[sks+05]\\	
MACS J0159.8$-$0849&29.95555 &$ -8.83303$&  0.4052&  17.82&  1.71& 104.64&   86& $  0.22\pm0.05$&R, relaxed[ars+08]\\		
A370             &  39.96969 &$ -1.57192$&  0.3732&  17.73&  2.18& 183.49&  139& $  0.07\pm0.06$&U, merger[omf98]\\
A383             &  42.01413 &$ -3.52923$&  0.1883&  15.94&  1.71& 100.85&   70& $  0.41\pm0.04$&R, relaxed[ars+08],cool[zfb+07]\\	
RX J0256.5+0006  &  44.12854 &    0.10091&  0.3665&  18.04&  1.58&  62.74&   48& $ -1.81\pm0.13$&U, merger[mnr+04]\\		    
A586             & 113.08453 &   31.63353&  0.1710&  14.76&  1.90& 160.23&  126& $  0.43\pm0.02$&R, relaxed[cls+05],cool[bfs+05]\\
A611             & 120.23675 &   36.05654&  0.2873&  16.66&  1.79& 107.64&   83& $  0.11\pm0.06$&R, relaxed[ars+08]\\
A644             & 124.35672 &$ -7.51257$&  0.0705&  14.01&  1.62&  77.30&   48& $ -0.49\pm0.07$&U, merger[bhs05]\\
RXC J0821.8+0112 & 125.46104 &    1.19709&  0.0871&  14.98&  1.58&  63.22&   44& $ -0.47\pm0.10$&U, merger[wbs+13]\\
A665             & 127.73873 &   65.84197&  0.1830&  16.25&  2.07& 208.30&  178& $ -0.26\pm0.10$&U, merger[mv01],radio[fgg+12]\\   
MS0839.8+2938    & 130.73311 &   29.45750&  0.1937&  16.34&  1.37&  52.05&   46& $  0.37\pm0.03$&R, cool[hmd10]\\
A697             & 130.73982 &   36.36646&  0.2823&  17.53&  1.44&  51.64&   64& $ -0.22\pm0.06$&U, merger (line of sight)[gbb06],radio[fgg+12]\\    
Z1953            & 132.53297 &   36.07046&  0.3764&  18.33&  2.02& 113.13&  111& $ -0.36\pm0.06$&U, merger[me12]\\    
A750             & 137.30312 &   10.97475&  0.1763&  15.56&  2.00& 148.73&  151& $  0.19\pm0.05$&R, cool[bfs+05]\\
A746             & 137.37737 &   51.51784&  0.2320&  17.91&  1.68&  85.48&   81& $ -2.49\pm0.10$&U, radio[fgg+12]\\   
IRAS 09104+4109  & 138.43956 &   40.94117&  0.4408&  18.17&  1.50&  58.35&   40& $  0.40\pm0.03$&R, cool[all00]\\    
A773             & 139.47258 &   51.72710&  0.2173&  16.08&  2.13& 149.40&  134& $ -0.12\pm0.07$&U, merger[bbg+07],radio[fgg+12]\\   
A781             & 140.20116 &   30.47176&  0.2881&  17.04&  1.85& 172.82&  127& $ -0.61\pm0.07$&U, merger[shw+08],radio[fgg+12]\\   
A800             & 142.15269 &   37.78979&  0.2450&  17.12&  1.94& 106.21&   79& $ -0.21\pm0.07$&U, radio[gff+12]\\   
A851             & 145.73941 &   46.98050&  0.4069&  18.43&  1.63&  86.69&   69& $ -1.15\pm0.05$&U, merger[dsc03],radio[fgg+12]\\
Z2701            & 148.20488 &   51.88484&  0.2152&  16.09&  1.64&  69.09&   52& $  0.25\pm0.03$&R, cool[bfs+05]\\
A910             & 150.76340 &   67.17450&  0.2304&  16.28&  2.19& 214.95&  116& $ -0.45\pm0.08$&U, radio[gff+12]\\   
MACS J1006.9+3200& 151.72778 &   32.02545&  0.3983&  17.85&  1.67&  68.02&   59& $ -0.59\pm0.10$&U, merger[me12]\\   
A963             & 154.26514 &   39.04706&  0.2056&  15.93&  1.69&  60.56&   66& $  0.05\pm0.04$&R, relaxed[ars+08],cool[all00]\\   
A959             & 154.39302 &   59.56106&  0.2880&  17.76&  1.87& 104.91&   80& $ -1.12\pm0.06$&U, merger[bbg09]\\   
Z3146            & 155.91515 &    4.18629&  0.2898&  16.84&  1.41&  60.82&   47& $  0.39\pm0.02$&R, relaxed[ars+08],cool[all00]\\   
A1201            & 168.22708 &   13.43584&  0.1681&  15.23&  1.71&  95.73&   78& $ -0.16\pm0.10$&U, merger[onc+09]\\   
MACS J1115.8+0129& 168.96626 &    1.49863&  0.3520&  17.68&  1.51&  54.71&   51& $  0.42\pm0.04$&R, relaxed[ars+08]\\
A1240            & 170.90714 &   43.05779&  0.1957&  16.56&  1.54&  84.65&   66& $ -0.52\pm0.12$&U, merger[bgb+09],radio[fgg+12]\\    
A1278            & 172.53789 &   20.51503&  0.1300&  15.44&  1.86& 108.42&   90& $ -0.29\pm0.10$&U, merger[mol+03]\\   
A1351            & 175.60326 &   58.53488&  0.3229&  17.07&  2.21& 130.13&   94& $ -1.23\pm0.15$&U, merger[hsd09],radio[fgg+12]\\   
A1413            & 178.82501 &   23.40494&  0.1429&  14.52&  1.82& 100.78&   90& $  0.36\pm0.02$&R, relaxed[ars+08],cool[all00]\\    
A1423            & 179.32217 &   33.61094&  0.2130&  16.13&  1.62&  79.82&   77& $  0.16\pm0.03$&R, cool[bfs+05]\\
A1550            & 187.26057 &   47.62238&  0.2540&  17.03&  2.06& 136.44&  111& $ -0.42\pm0.09$&U, radio[gff+12]\\   
A1560            & 188.47086 &   15.19464&  0.2835&  17.47&  1.36&  68.35&   54& $ -0.74\pm0.09$&U, merger[uc82]\\    
A1589            & 190.32281 &   18.57457&  0.0704&  13.96&  1.81& 101.94&   71& $ -0.61\pm0.13$&U, merger[bro+94]\\    
A1612            & 191.97337 &$ -2.80733$&  0.1819&  16.33&  1.91& 111.88&   81& $ -1.71\pm0.12$&U, radio[fgg+12]\\    
A1650            & 194.67290 &$ -1.76146$&  0.0839&  14.08&  1.51&  75.02&   51& $  0.21\pm0.03$&R, cool[dvo+05]\\
A1682            & 196.70831 &   46.55927&  0.2257&  16.16&  2.04& 126.85&  115& $ -0.87\pm0.06$&U, merger[mol+03],radio[vgd+08]\\    
ZwCl1305.4+2941  & 196.95512 &   29.43006&  0.2405&  16.14&  1.41&  55.94&   38& $  0.48\pm0.03$&R, relaxed[gtv+08]\\
A1689            & 197.87291 &$ -1.34109$&  0.1828&  15.69&  2.28& 172.53&  156& $  0.47\pm0.05$&U, merger[am04],radio[fgg+12],cool[all00]\\
A1704            & 198.60248 &   64.57538&  0.2191&  16.50&  1.63&  88.00&   64& $  0.00\pm0.06$&R, cool[all00]\\    
A1703            & 198.77182 &   51.81738&  0.2836&  16.76&  1.96& 130.12&  100& $  0.16\pm0.04$&R, relaxed[rpl+09]\\ 
A1750            & 202.79594 &$ -1.72730$&  0.0835&  14.47&  1.85& 134.22&   83& $ -0.09\pm0.04$&U, merger[dfj+01,bps+04]\\   
A1758            & 203.16005 &   50.56000&  0.2794&  17.33&  1.98& 155.02&  129& $ -0.70\pm0.08$&U, merger[dk04],radio[fgg+12]\\   
A1763            & 203.83372 &   41.00115&  0.2278&  16.37&  1.96& 155.64&  146& $  0.13\pm0.05$&U, merger[st08]\\   
A1795            & 207.21877 &   26.59293&  0.0633&  13.67&  1.50&  51.87&   51& $  0.36\pm0.05$&R, relaxed[ars+08],cool[all00]\\   
MS1358.4+6245    & 209.96069 &   62.51819&  0.3271&  17.77&  1.72&  85.96&   75& $ -0.69\pm0.10$&U, merger[fff+98],cool[all00]\\   
A1835            & 210.25864 &    2.87847&  0.2520&  16.06&  2.22& 192.81&  178& $  0.56\pm0.02$&R, relaxed[ars+08],cool[all00]\\
A1904            & 215.54253 &   48.57074&  0.0710&  13.77&  1.75& 102.75&   68& $ -0.49\pm0.12$&U, merger[blw+96]\\			    
A1914            & 216.48611 &   37.81646&  0.1700&  15.62&  2.03& 135.55&  117& $ -0.36\pm0.10$&U, merger[gmv+04],radio[fgg+12]\\    
MaxBCG J217.9+13.5&217.95869 &   13.53471&  0.1599&  15.40&  1.44&  50.58&   32& $ -0.15\pm0.05$&U, radio[nhv12]\\
A1942            & 219.59111 &    3.67035&  0.2247&  16.25&  1.55&  81.55&   71& $  0.24\pm0.04$&R, relaxed[cpl+08]\\
A1995            & 223.23949 &   58.04876&  0.3212&  17.71&  1.80&  83.59&   83& $ -0.09\pm0.07$&U, radio[fgg+12]\\		    
A1991            & 223.63121 &   18.64232&  0.0592&  13.61&  1.59&  68.09&   51& $ -0.41\pm0.05$&R, relaxed[vmm+05],cool[wok+10]\\
Z7160            & 224.31294 &   22.34289&  0.2576&  16.37&  1.66&  78.77&   54& $  0.22\pm0.04$&R, cool[all00]\\		    
RXC J1504.1$-$0248&226.03130 &$ -2.80460$&  0.2169&  16.27&  1.68&  80.30&   66& $  0.32\pm0.04$&R, relaxed[ars+08],cool[wok+10]\\
A2034            & 227.54880 &   33.48647&  0.1116&  14.75&  1.89& 116.63&   89& $ -0.25\pm0.06$&U, merger[ksm03],radio[fgg+12]\\   
A2029            & 227.73376 &    5.74478&  0.0779&  13.36&  1.71&  87.42&   91& $  0.40\pm0.03$&R, relaxed[ars+08],cool[all00]\\
A2048            & 228.80881 &    4.38622&  0.0950&  15.21&  1.64&  71.28&   72& $ -0.74\pm0.07$&U, radio[fgg+12]\\

\hline
\end{tabular}}
\end{center}
\label{tab1}
\end{table*} 
                                                                                                                                    
\addtocounter{table}{-1}                                                                                                             
\begin{table*}
\caption[]{{\it continued}}
\begin{center}
\resizebox{\textwidth}{!}{
\begin{tabular}{@{}lrrrccrrrl@{}}

\hline
\mc{1}{l}{Name}    &\mc{1}{c}{R.A.}       &\mc{1}{c}{Dec.}   &\mc{1}{c}{$z$}      & \mc{1}{c}{$r_{\rm BCG}$}& 
\mc{1}{c}{$r_{200}$}&\mc{1}{c}{$R_{L\ast}$}&\mc{1}{c}{$N_{200}$}&\mc{1}{c}{$\Gamma$} & \mc{1}{l}{Comments and reference}\\
                   &\mc{1}{c}{(deg)}      &\mc{1}{c}{(deg)}   &                    &                        &
\mc{1}{c}{(Mpc)}    &                     &                    &                    &                    \\
(1) & \mc{1}{c}{(2)} & \mc{1}{c}{(3)} & \mc{1}{c}{(4)} & (5) & (6) & \mc{1}{c}{(7)} & (8) & \mc{1}{c}{(9)}& (10) \\
\hline

RXC J1516.5$-$0056&229.24202 & $-1.11080$&  0.1186&  15.29&  1.29&  61.43&   49& $ -1.24\pm0.10$&U, merger[wbs+13]\\
A2061            & 230.33575 &   30.67093&  0.0788&  13.99&  1.86&  87.42&   90& $ -0.58\pm0.11$&U, merger[mbz+04],radio[fgg+12]\\
A2065            & 230.60008 &   27.71437&  0.0726&  14.84&  1.90& 102.58&  113& $ -0.53\pm0.07$&U, merger[msv99]\\	    
A2069            & 231.03093 &   29.88896&  0.1135&  14.89&  1.80& 114.07&  114& $ -0.26\pm0.04$&U, merger[onc+09]\\		    
RX J1532.9+3021  & 233.22408 &   30.34984&  0.3621&  17.34&  1.60&  74.22&   59& $  0.28\pm0.04$&R, relaxed[ars+08],cool[bfs+05]\\	
A2111            & 234.91873 &   34.42424&  0.2280&  16.95&  2.03& 134.29&  131& $ -0.04\pm0.05$&U, merger[wul97]\\
A2142            & 239.58333 &   27.23341&  0.0910&  14.40&  2.03& 140.29&  154& $ -0.31\pm0.06$&U, merger[mpn+00],cool[all00]\\   
MS1621.5+2640    & 245.89771 &   26.57057&  0.4270&  18.69&  1.61&  77.81&   67& $ -0.59\pm0.04$&U, merger[sqf+10]\\
A2219            & 250.08255 &   46.71153&  0.2244&  16.54&  1.85& 148.59&  148& $ -0.24\pm0.06$&U, merger[bgb+04],radio[fgg+12]\\   
A2244            & 255.67706 &   34.06000&  0.0994&  14.02&  1.76&  65.80&   81& $  0.48\pm0.03$&R, cool[dvo+05]\\
A2256            & 256.11325 &   78.64049&  0.0594&  13.28&  1.71&  67.64&   42& $ -0.17\pm0.08$&U, merger[smm+02],radio[fgg+12]\\    
A2255            & 258.11984 &   64.06083&  0.0808&  13.97&  2.14& 139.88&  111& $ -1.02\pm0.10$&U, merger[brp+95],radio[fgg+12]\\   
A2259            & 260.04019 &   27.66889&  0.1640&  15.26&  1.36&  53.46&   45& $  0.36\pm0.03$&R, cool[bfs+05]\\
RX J1720.1+2638  & 260.04184 &   26.62557&  0.1601&  15.51&  1.54&  70.83&   60& $  0.33\pm0.03$&R, relaxed[onc+09],cool[bfs+05]\\
MACS J1720.3+3536 &260.06989 &   35.60736&  0.3913&  17.85&  1.62&  79.19&   62& $  0.02\pm0.05$&R, relaxed[ars+08],cool[wok+10]\\	
A2261            & 260.61325 &   32.13257&  0.2233&  15.26&  2.18& 219.92&  176& $  0.25\pm0.03$&R, cool[all00]\\			    
MACS J1731.6+2252& 262.91638 &   22.86628&  0.3660&  17.72&  2.07& 123.13&   88& $ -0.41\pm0.10$&U, merger[me12],radio[bbv+12]\\		    
RX J2129.6+0006  & 322.41647 &    0.08921&  0.2339&  16.32&  1.65&  84.41&   61& $  0.42\pm0.04$&R, relaxed[ars+08],cool[bfs+05]\\	
A2356            & 323.94287 &    0.11587&  0.1175&  15.91&  1.49&  50.95&   40& $ -1.90\pm0.09$&U, merger[uc82]\\
A2390            & 328.40347 &   17.69548&  0.2302&  16.66&  1.87& 104.12&  107& $  0.04\pm0.06$&R, relaxed[ars+08],cool[all00]\\   
RXC J2157.4$-$0747&329.25717 &$ -7.83959$&  0.0583&  13.97&  1.54&  63.38&   45& $ -1.36\pm0.11$&U, merger[wbs+13]\\
A2440            & 335.98724 &$ -1.58326$&  0.0900&  14.38&  1.88&  78.46&   52& $ -0.44\pm0.10$&U, merger[msb+11]\\   
A2443            & 336.53302 &   17.35651&  0.1070&  14.64&  1.77&  97.42&   69& $ -1.20\pm0.10$&U, radio[cc11]\\					    
MACS J2228.5+2036& 337.14050 &   20.62120&  0.4120&  18.64&  1.95& 153.70&  109& $ -0.89\pm0.12$&U, merger[me12]\\					    
MACS J2243.3$-$0935&340.83633&$ -9.58858$&  0.4393&  19.76&  1.83&  93.70&   93& $ -1.53\pm0.07$&U, merger[me12]\\					    
A2537            & 347.09262 &$ -2.19217$&  0.2950&  16.79&  2.13& 107.27&  101& $  0.06\pm0.05$&R, relaxed[ars+08]\\
A2626            & 354.12756 &   21.14734&  0.0546&  13.17&  1.56&  66.98&   47& $  0.07\pm0.05$&R, relaxed[wbs+13]\\
A2631            & 354.41553 &    0.27137&  0.2772&  17.24&  1.75&  97.17&   91& $ -0.02\pm0.10$&U, merger[me12]\\	    
ZwCl2341.1+0000  & 355.94806 &    0.25666&  0.2673&  17.99&  1.66& 104.48&   75& $ -0.56\pm0.08$&U, merger[ak13],radio[nhv+12]\\	    	    

\hline
\end{tabular}}
\end{center}
{Note. 
Column (1): cluster name. 
Column (2)--(4): R.A., Dec. (J2000) and redshift of cluster. 
Column (5): $r$-band magnitude of BCG.
Column (6): cluster radius, $r_{200}$ (Mpc). 
Column (7): cluster richness.
Column (8): number of member galaxies within $r_{200}$.
Column (9): relaxation parameter.
Comments in column (10): `R/U' refers to classification (relaxed/unrelaxed) of clusters in this work.  
`merger' means the cluster shows merging signatures; `radio' means the cluster has radio 
halo/relic; `cool' means the cluster has a cool core in X-ray. Reference for `merger', `radio' and `cool' in column (10):  
ak13 \citep{ak13}; all00 \citep{all00}; am04 \citep{am04}; ars+08 \citep{ars+08}; 
bbg+07 \citep{bbg+07}; bbg09 \citep{bbg09}; bbv+12 \citep{bbv+12}; 
bfs+05 \citep{bfs+05}; bgb+04 \citep{bgb+04}; bgb+09 \citep{bgb+09}; bhs05 \citep{bhs05}; blw+96 \citep{blw+96}; bps+04 \citep{bps+04}; 
bro+94 \citep{bro+94}; brp+95 \citep{brp+95}; cc11 \citep{cc11}; cls+05 \citep{cls+05}; 
cmk+02 \citep{cmk+02}; cpl+08 \citep{cpl+08}; dfj+01 \citep{dfj+01}; dk04 \citep{dk04}; dsc03 \citep{dsc03}; 
dvo+05 \citep{dvo+05}; fbb+81 \citep{fbb+81}; fff+98 \citep{fff+98}; fgg+12 \citep{fgg+12}; 
gbb06 \citep{gbb06}; gff+12 \citep{gff+12}; gmv+04 \citep{gmv+04}; gtv+08 \citep{gtv+08}; 
hmd10 \citep{hmd10}; hsd09 \citep{hsd09}; jsc+08 \citep{jsc+08}; ksm03 \citep{ksm03}; 
ksr02 \citep{ksr02}; mbz+04 \citep{mbz+04}; me12 \citep{me12}; mnr+04 \citep{mnr+04}; 
mol+03 \citep{mol+03}; mpn+00 \citep{mpn+00}; msb+11 \citep{msb+11}; msv99 \citep{msv99}; 
mv01 \citep{mv01}; nhv+12 \citep{nhv+12}; omf98 \citep{omf98}; onc+09 \citep{onc+09}; 
rpl+09 \citep{rpl+09}; shw+08 \citep{shw+08}; sks+05 \citep{sks+05}; smm+02 \citep{smm+02}; 
sqf+10 \citep{sqf+10}; st08 \citep{st08}; uc82 \citep{uc82}; vgd+08 \citep{vgd+08}; vmm+05 \citep{vmm+05}; 
wbs+13 \citep{wbs+13};
wok+10 \citep{wok+10}; wul97 \citep{wul97}; zfb+07 \citep{zfb+07}.
}
\end{table*}

\section{Test sample of galaxy clusters with known dynamical states}

We collect the test sample of clusters with known dynamical states to
verify the method for quantification of dynamical states of clusters
by using optical photometric data and to find optimized parameters.

\subsection{Sample collection}

Galaxy clusters have been broadly classified as dynamically
`relaxed' and `unrelaxed'.
In literature, a few hundreds of clusters have their dynamical states
so classified according to their X-ray characteristics or redshift
distributions of member galaxies
\citep[e.g.,][]{cmk+02,vmm+05,cls+05,ars+08,me12}, which we have
collected into the parent sample. We also collect the clusters based
on X-ray and radio observations. Most clusters with a cool core in
X-ray are known to be relaxed \citep{esf92,afj+01,bfs+05,vmm+05}.
Therefore, clusters are collected as relaxed clusters if they have a
central cooling time less than 1 Gyr \citep{hmd10,wok+10} or a cooling
time less than 10 Gyr within a radius greater than 50 kpc
\citep{all00,bfs+05}, except for a few cool-core clusters with
distinct disturbed X-ray morphologies \citep[e.g., A85 and
  A115,][]{ksr02, dnf05, gk05, bfs+05} which we classify as unrelaxed
clusters.
Clusters with radio halos or radio relics are exclusively unrelaxed
systems \citep[e.g.,][]{gmv+04,bbg+07,bgb+09,bbg09,ceg+10}. As long as
radio halo/relics were detected \citep{vgd+08,cc11,fgg+12,nhv+12}, we
collect the clusters into the parent sample as unrelaxed
clusters. 

From the parent sample with properly known dynamical states for
relaxed and unrelaxed clusters, we choose the test sample of clusters
only in the sky area of SDSS-III so that we can get their optical data
for member galaxies. To ensure the complete detection of luminous
member galaxies \citep{whl12}, we further limit the cluster sample to
be in the redshift range of $0.05<z\lesssim0.42$. We also set the
threshold of cluster richness (defined below) $R_{L\ast} \ge 50$ so
that the test clusters have enough recognized member galaxies. With
these selection criteria, we get the test sample of 98 clusters with
known dynamical states, including 35 relaxed and 63 unrelaxed
clusters, as listed in Table~\ref{tab1}. Among the 35 relaxed
clusters, 24 have been classified in literature already (see
references in Table~\ref{tab1}), 11 clusters are included because of
their cool cores shown in X-ray. Among 63 unrelaxed clusters, 54
clusters have their classification made previously in literature and 9
clusters are included because of their radio halos or relics.

\subsection{Member galaxies discrimination and richness estimation}
\label{radrich}

For the test sample of 98 clusters and also the work sample of
2092 clusters for Section~\ref{dis_gamma}, we use the photometric data
of the SDSS Data Release 8\footnote{http://www.sdss3.org/dr8/}
\citep[SDSS DR8,][]{dr8+11} to discriminate luminous member galaxies
and to estimate cluster richness and radius which will be used in the
quantification of cluster dynamical state. If any objects in the SDSS
DR8 have deblending problems and saturated, they are discarded using
the flags\footnote{(flags \& 0x20) = 0 and (flags \& 0x80000) = 0 and
  ((flags \& 0x400000000000) = 0 or psfmagerr$_r <= 0.20$) and ((flags
  \& 0x40000) = 0)}.

In our work, the location of the first brightest cluster galaxy (BCGs)
identified from optical photometric data is taken as the cluster
center. This is reasonable because for clusters with X-ray emission,
\citet{whl09} showed that most of the first BCGs have a projected
separation less than 0.2 Mpc from the X-ray peaks. 

Luminous member galaxy candidates of a cluster are selected according
to photometric redshifts (hereafter photo-$z$s) and the projected
separation from the center. The uncertainties of photo-$z$s are
$\sigma_z\sim$0.025--0.030 in the redshift range of $z<0.45$
\citep{whl12}. We assume that the uncertainty increases with redshift
in the form of $\sigma_z=\sigma_0(1+z)$ for all galaxies. Member
galaxy candidates brighter than $M^{\rm e}_r(z)\le-20.5$ are selected
from SDSS data within a photo-$z$ slice of $z\pm 0.04(1+z)$. Here,
$M^{\rm e}_r$ is the evolution-corrected absolute magnitude in the
$r$ band, $M^{\rm e}_r(z)=M_r(z)+Qz$, where we adopt a passive
evolution of $Q=1.62$ \citep{bhb+03}. The completeness of such
selected member galaxies is more than 90\% for rich clusters
\citep{whl09}. Objects with a large photo-$z$ error, $z{\rm
  Err}>0.08(1+z)$, are discarded for member galaxy candidates because
they suffer bad photometry or contamination of stars. In this thick
redshift slice, the relatively large uncertainty of photo-$z$ leads to
member galaxy candidates contaminated by foreground and background
galaxies. To further diminish the contamination and incompleteness, we
complement the photometric redshift with the spectroscopic
measurements of the SDSS DR9 \citep{dr9+12}. Member galaxy candidates
within the projected separation $<$2\,Mpc from the cluster center are
discarded from the list if they have a velocity difference $\Delta
v>2500$ km~s$^{-1}$ from cluster redshift. For completeness, we
include the missing galaxies in the photo-$z$ data if they have a
velocity difference $\Delta v\le2500$ km~s$^{-1}$ from the cluster
redshift and a separation of $<$2\,Mpc from the cluster center.

\begin{figure}
\centering
\includegraphics[width = 8.cm]{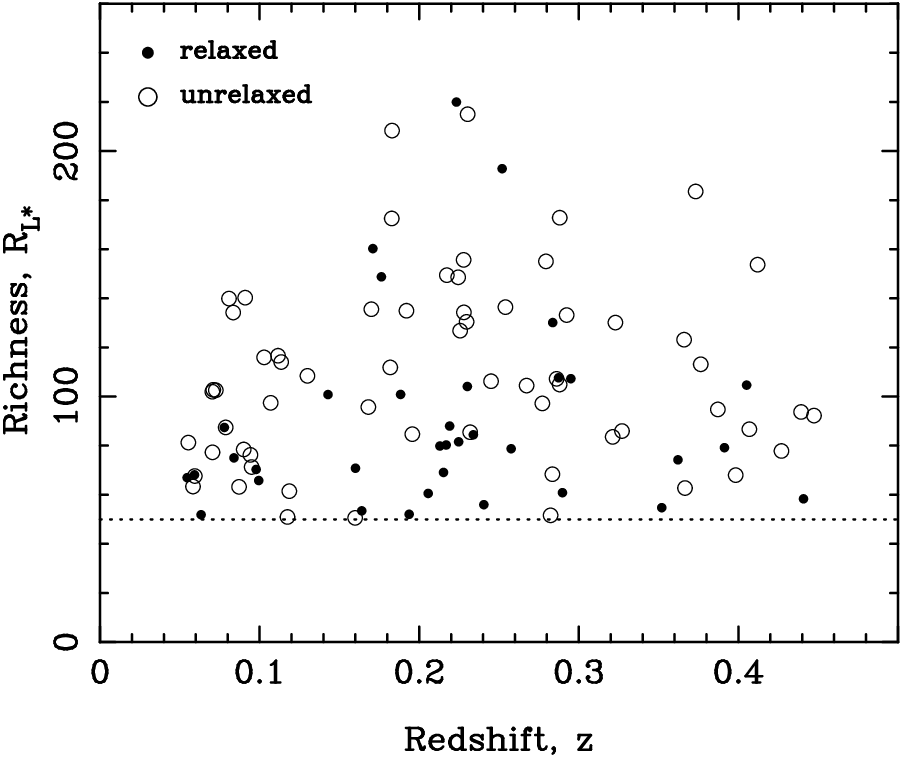}
\caption{Richness distribution of the test sample of 98 clusters with
  known dynamical states in the redshift range of
  $0.05<z\lesssim0.42$. 35 relaxed clusters are plotted as black dots,
  and 63 unrelaxed clusters are plotted as open circles. The dotted
  line indicates $R_{L\ast}=50$.}
\label{zrich}
\end{figure}

Cluster radius and richness are estimated from this list of member
galaxy candidates by using the procedure of \citet{whl12}. For each
cluster, we first get the sum of the $r$ band luminosities of member
galaxies within a photo-$z$ slice of $z\pm 0.04(1+z)$ and within 1 Mpc
from the cluster center, $L_{\rm 1Mpc}$, with a local background
subtracted\footnote{To estimate local background, we follow the method
  of \citet{pbb+04} and divide the annuals between 2 and 4 Mpc for each
  cluster around the BCG into 48 sections with an equal area. Within
  the same magnitude and photo-$z$ range, we calculate the $r$-band
  total luminosity in each section and estimate the mean value and its
  root mean square. The regions with luminosities larger than
  3$\,\sigma$ are discarded and the mean is recalculated and taken as
  the local background.}.
We then relate $L_{\rm 1Mpc}$ to cluster radius, $r_{200}$, i.e., the
radius within which the mean density of a cluster is 200 times of the
critical density of the universe, by using the relation \citep{whl12}
\begin{equation}
\log r_{200}=-0.57+0.44\log( L_{\rm 1Mpc}/L^{\ast} ),
\label{r200}
\end{equation}
where $r_{200}$ is in units of Mpc, $L^{\ast}$ is the evolved
characteristic luminosity of galaxies in the $r$ band, defined as
$L^{\ast}(z)=L^{\ast}(z=0)10^{0.4Qz}$ \citep{bhb+03}.
The $r$-band total luminosity $L_{200}$ are then obtained from the sum
of $r$-band luminosity of cluster members within the radius of
$r_{200}$. The cluster richness is thus defined as
$R_{L\ast}=L_{200}/L^{\ast}$. For the test sample, we include rich
clusters with $R_{L\ast}\ge 50$, which corresponds to
$M_{200}\ge3.15\times10^{14}~M_{\odot}$ according to equation~2 of
\citet{whl12}. The radius and richness for 98 test clusters are listed
in Table~\ref{tab1}, and their distribution across redshift and
richness is plotted in Fig.~\ref{zrich}.

\section{Quantification of cluster dynamical states from photometric data}

We quantify substructures of the two-dimensional optical maps for
dynamical states of clusters by using the two-dimensional brightness
distribution of cluster member galaxies. Because more luminous member
galaxies can trace more mass, all member galaxies selected above for
each cluster are smoothed and weighted by their luminosities with a
Gaussian kernel to get the brightness distribution of a cluster in the
sky plane. We then calculate three quantities defined in the following
subsections: the asymmetry factor, $\alpha$, the ridge flatness,
$\beta$, and the normalized deviation, $\delta$, to quantify the
dynamical states of clusters. Considering that the contamination rate
of member galaxies increases with cluster centric distance, we work on
the smoothed optical luminosity distribution within the central region
of clusters within $r_{500}=2/3\;r_{200}$ \citep{sks+03} for dynamical
state of clusters. The value of $r_{200}$ is estimated by using
Equation~(\ref{r200}).

\begin{figure*}
\centering
\resizebox{55mm}{!}{\includegraphics{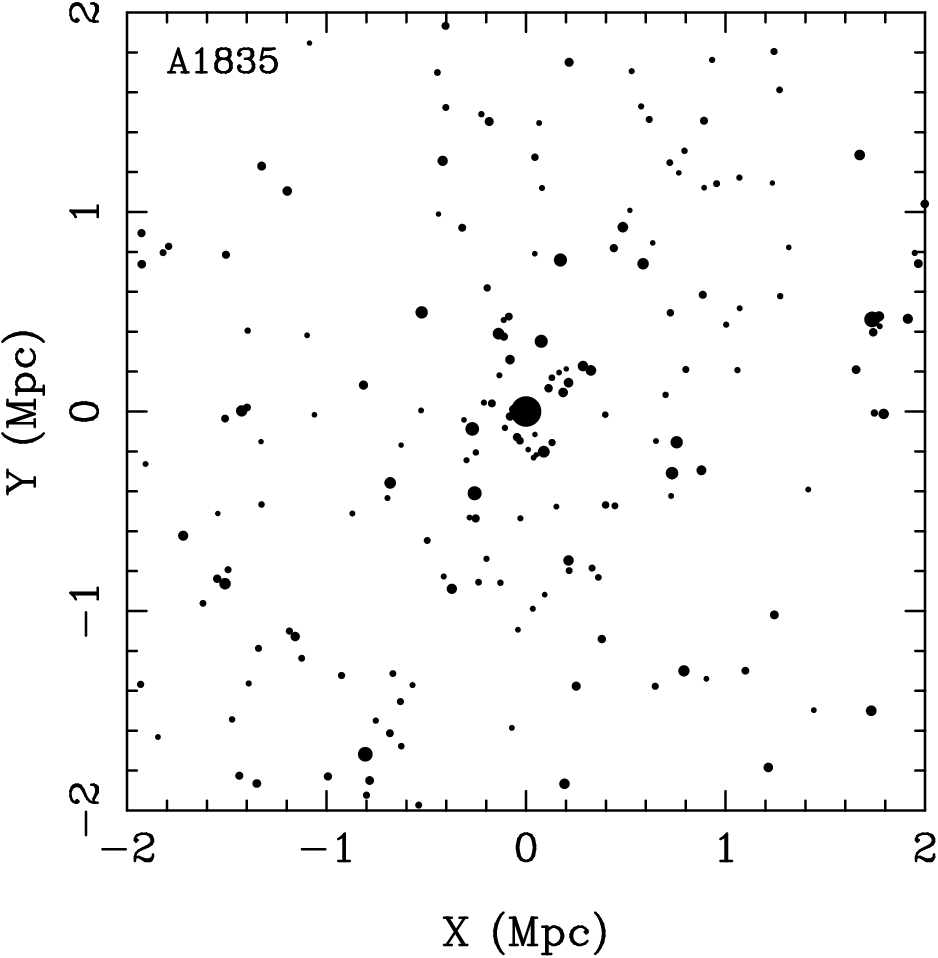}}~%
\resizebox{55mm}{!}{\includegraphics{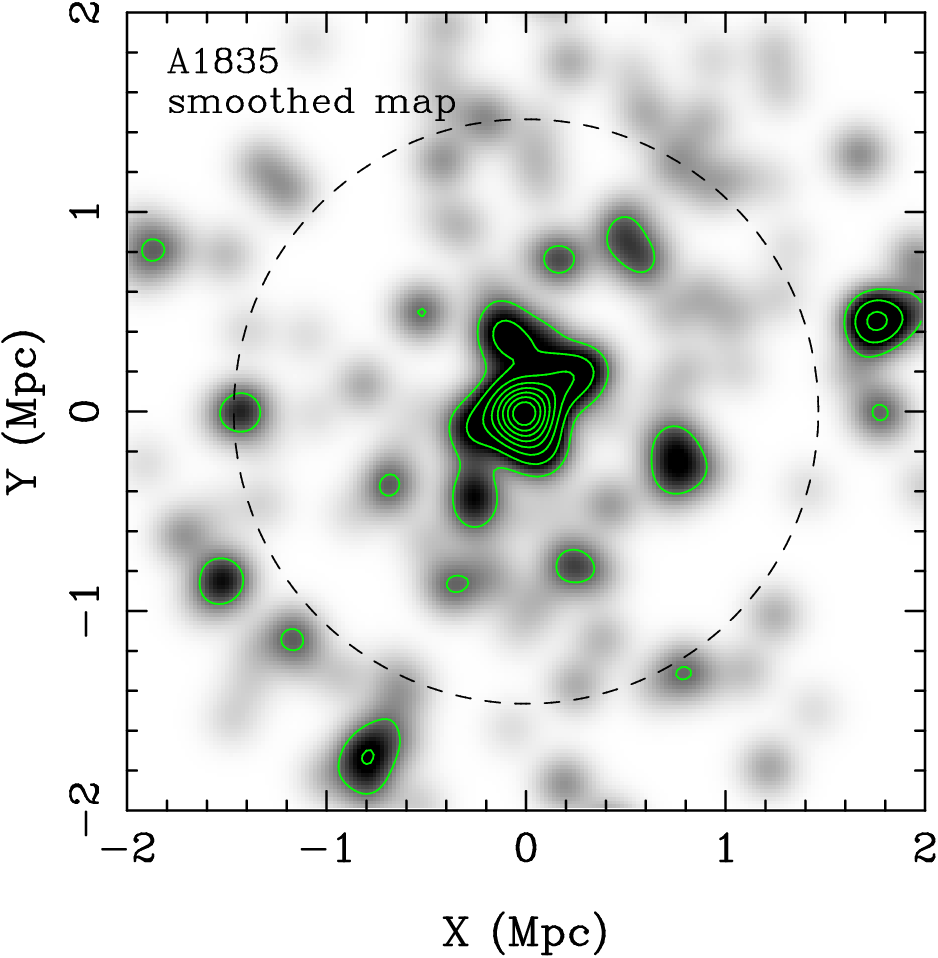}}~%
\resizebox{55mm}{!}{\includegraphics{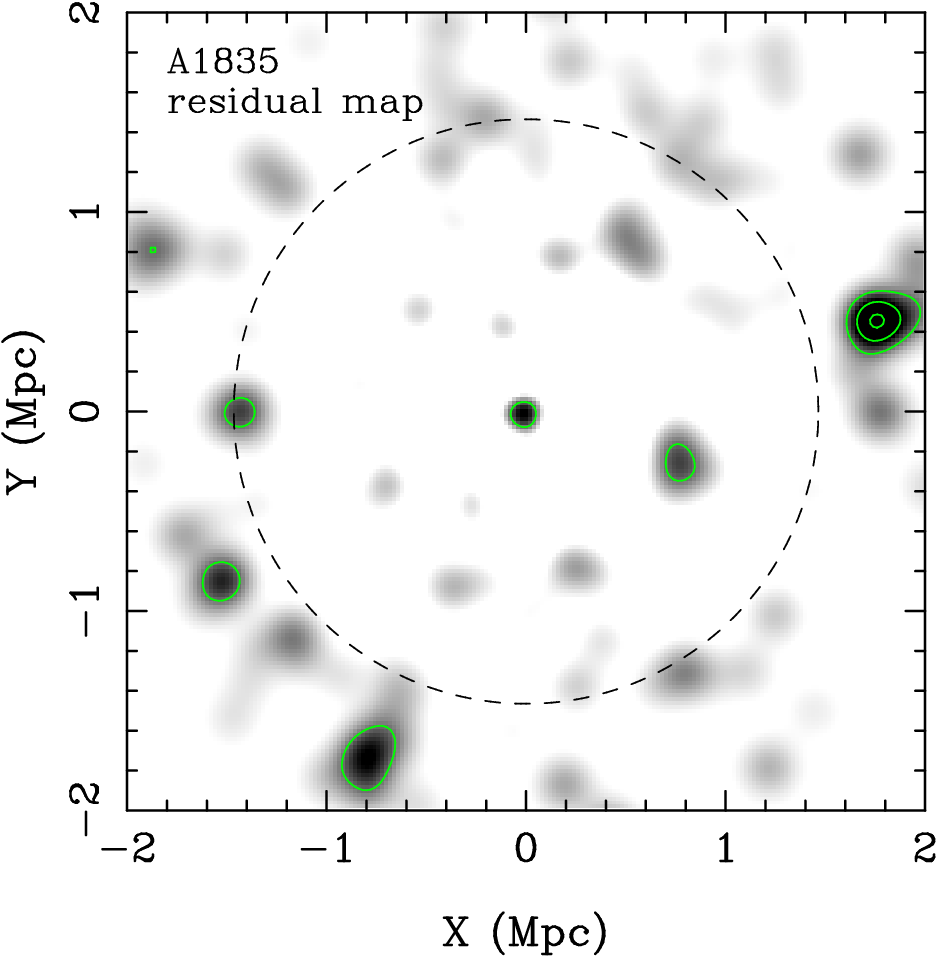}}\\[0.5mm]
\resizebox{55mm}{!}{\includegraphics{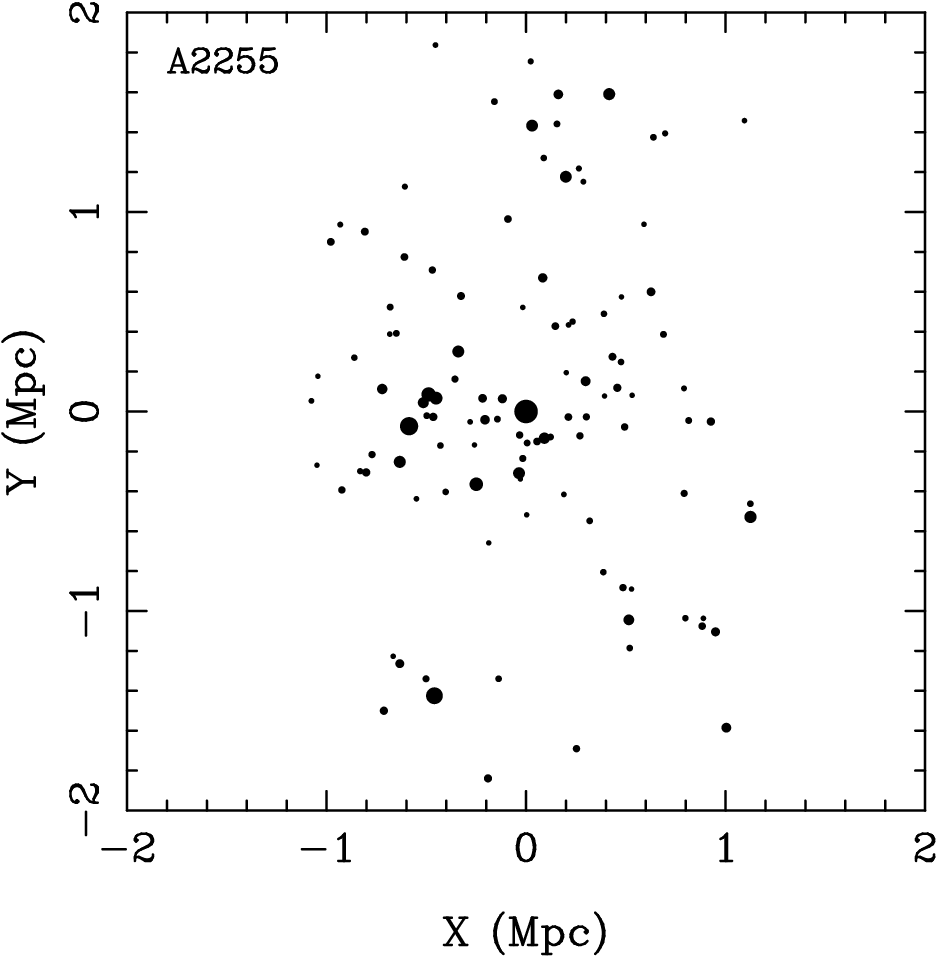}}~%
\resizebox{55mm}{!}{\includegraphics{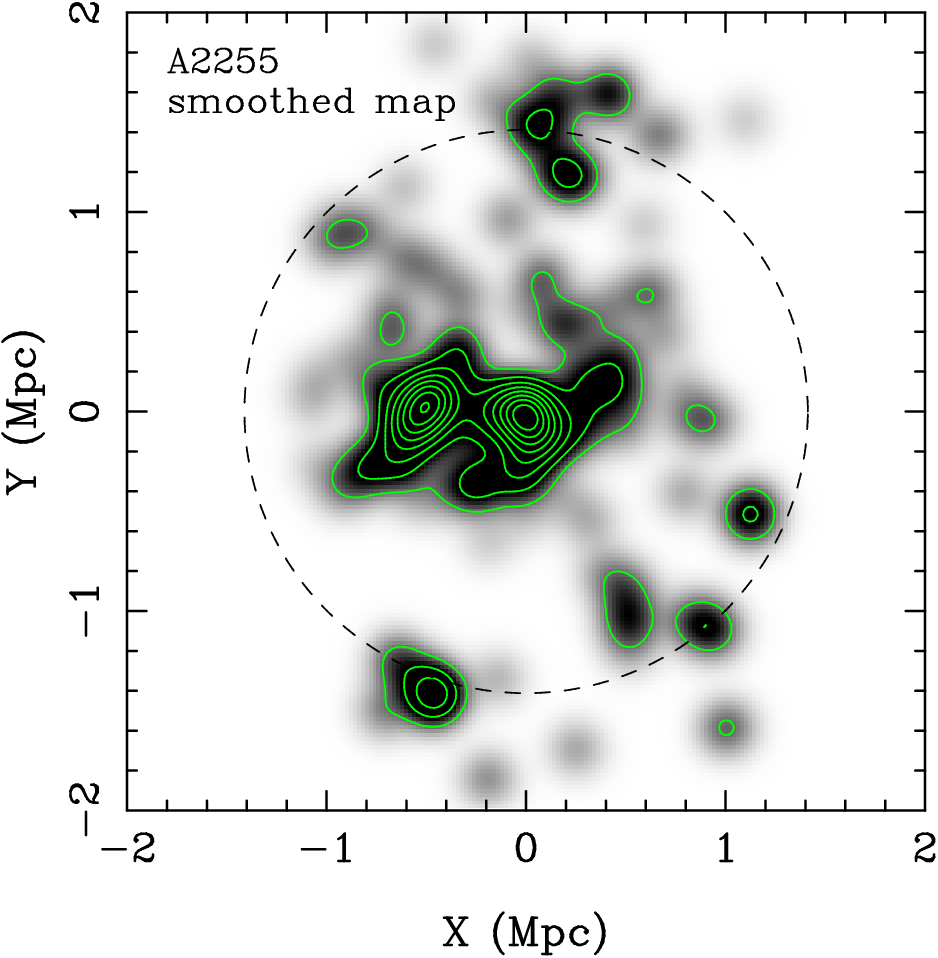}}~%
\resizebox{55mm}{!}{\includegraphics{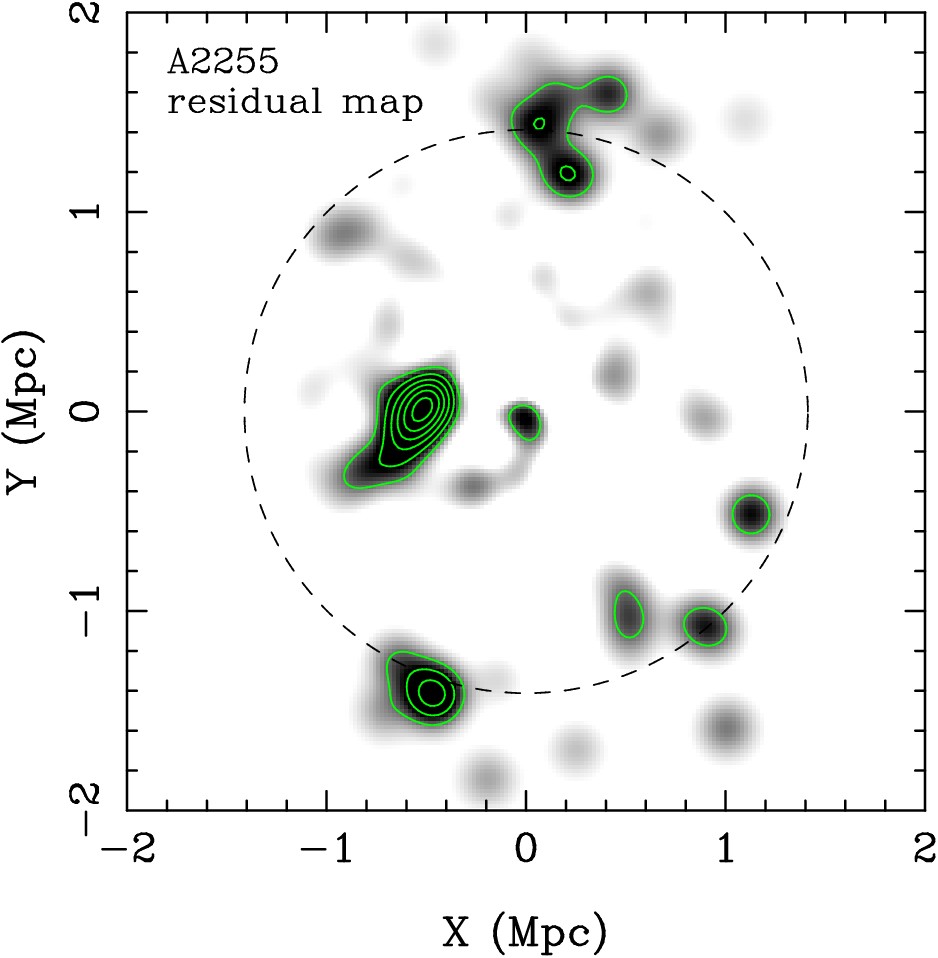}}
\caption{Projected distributions of member galaxies for a relaxed
  cluster, A1835 (upper panels), and a merging cluster, A2255 (lower
  panels). In the left panels, the black dots show the positions and
  the weight of galaxy luminosity ($L_r$) with the symbol size scaled
  by $\sqrt{L_r/L^{\ast}}$. The middle panels are the smoothed
  brightness maps of member galaxies convolved by a Gaussian function
  in Equation~(\ref{gauss}) with $\sigma=0.05r_{200}$ . The right
  panels are the residual maps after the smoothed map being subtracted by
  the best-fitting two-dimensional King model. The dashed circles indicate the
  regions with a radius of $r_{500}=2/3\;r_{200}$.
}
\label{exam1}
\end{figure*}

\subsection{The smoothed optical map}

From SDSS-III, we get the $r$-band positions and luminosities of
member galaxies of a cluster within a photo-$z$ slice of $z\pm
0.04(1+z)$ and a radius of $r_{200}$. The first step to quantify the
dynamical state is to smooth the optical photometric data with a
two-dimensional Gaussian function. Positions (RA, Dec.) of galaxies
are transformed into Cartesian coordinates ($x$, $y$) centering on the
cluster center. In practice, the region of 4~Mpc $\times$ 4 Mpc is
divided into 200 pixels $\times$ 200 pixels, so that each pixel has a
linear size of 20 kpc. The optical luminosity within each pixel
($x_i$,$y_j$) can be calculated by convolving all member galaxies with
a Gaussian kernel,
\begin{equation}
I(x_i,x_j)=\sum_{k=1}^{N_{200}}L_kg(x_i-x_k,y_j-y_k,\sigma_k),
\end{equation}
where $x_k$ and $y_k$ are the coordinates of the $k$th member galaxy,
$L_k$ is the $r$-band luminosity of this galaxy in unit of $L^{\ast}$,
$N_{200}$ is the total number of member galaxies within a region of a
photo-$z$ slice of $z\pm 0.04(1+z)$ and a radius of $r_{200}$,
$g(x,y,\sigma)$ is a two-dimensional Gaussian function with a smooth
scale of $\sigma$,
\begin{equation}
g(x,y,\sigma)=\frac{1}{2\pi \sigma^2}\exp\Big(-\frac{x^2+y^2}{2\sigma^2}\Big).
\label{gauss}
\end{equation}
It has been shown that light-to-mass ratios vary with cluster centric
distance \citep{kbm04,mbu+10}. The luminosities of member galaxies at
different radii are therefore related to galaxy mass with various
light-to-mass ratios, so that they contribute to dynamical state with
different smooth scale. Here, we take a smooth scale in the form of
\begin{equation}
\frac{\sigma}{r_{200}}=(\sigma_a+\sigma_b\frac{r}{r_{200}}),
\label{sscale}
\end{equation}
where $\sigma_a$ and $\sigma_b$ are the two parameters for smooth
scale, $r/r_{200}$ is projected distance of a member galaxy from the
cluster center in unit of $r_{200}$. Richer clusters have larger
$r_{200}$ and are smoothed with larger scales. This ensures that our
calculations below are independent of cluster radius or richness. The
background is subtracted as described above to give a net
two-dimensional smoothed brightness map of a cluster. Two examples are
shown in Fig.~\ref{exam1} for the projected luminosity distributions
of member galaxies (left panels), and the smoothed maps (middle
panels) and the residual maps (right panels, after subtraction of the
best-fitting two-dimensional King model, see discussions below) for a
relaxed cluster, A1835, and a merging cluster, A2255. The values of
$\sigma_a$ and $\sigma_b$ will be optimized in following calculations
for dynamical states of clusters.

\subsection{Asymmetry factor, $\alpha$}

A relaxed cluster has a regular symmetrical morphology, while an
unrelaxed cluster shows many substructures. The asymmetry of galaxy
distribution can be used to quantify the substructure of clusters. For
example, as shown in the middle panel of Fig.~\ref{exam1}, the
relaxed cluster, A1835, has a much more symmetrical map than the
unrelaxed cluster, A2255. Here, we calculate an asymmetry factor of
the smoothed map of member galaxies. First, the `total fluctuation
power' of the map within a radius of $r_{500}=2/3\;r_{200}$ is
calculated as
\begin{equation}
S^2=\sum_{i,j}I^2(x_i,y_j).
\label{eq_s}
\end{equation}
Then, we get the `difference power' of the map for all symmetric
pixels,
\begin{equation}
\Delta^2=\sum_{i,j}[I(x_i,y_j)-I(-x_{i},-y_{j})]^2/2.
\end{equation}
The asymmetry factor of a cluster, $\alpha$, is defined as 
\begin{equation}
\alpha=\frac{\Delta^2}{S^2}.
\end{equation}
Here, $\alpha=0$ implies a very symmetric morphology of a cluster,
whereas $\alpha=1$ indicates a highly asymmetric morphology.

\begin{figure}
\centering
\includegraphics[width = 7.5cm]{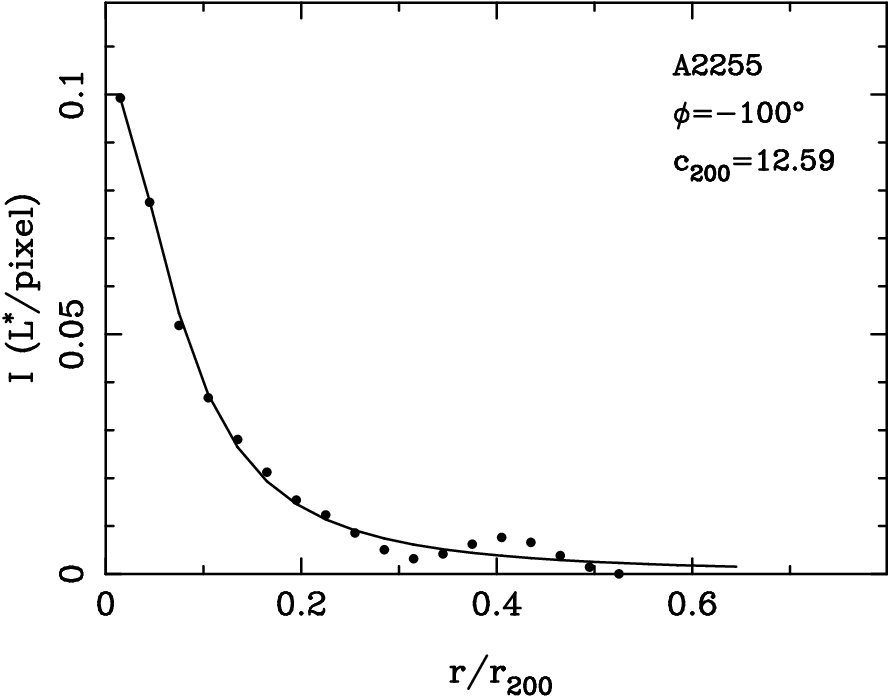}
\includegraphics[width = 7.5cm]{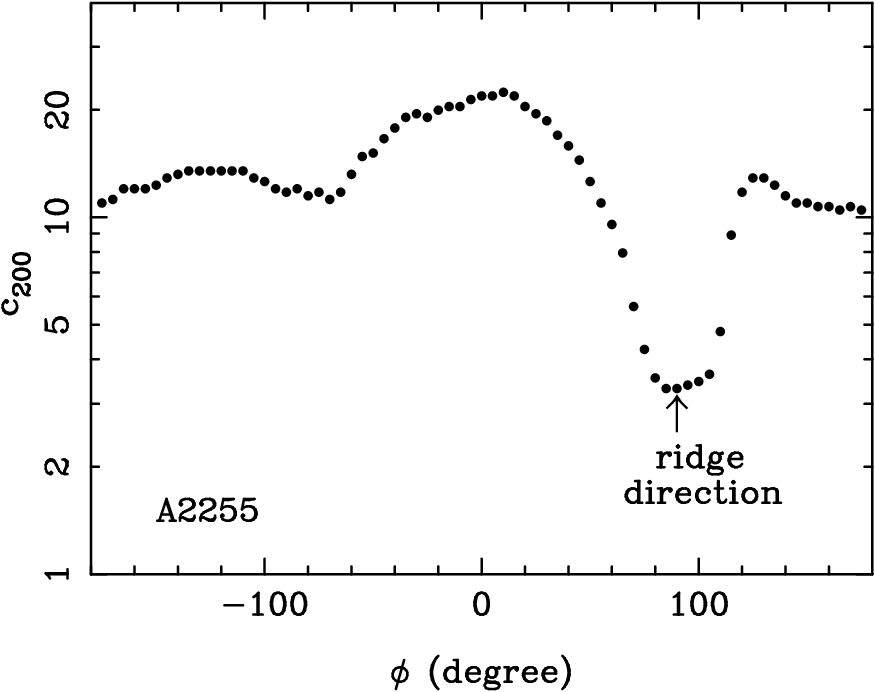}
\caption{An example for relative ridge steepness factor.  The
  upper panel shows the light profile at a direction and
  one-dimensional King model. The lower panel shows variations of
  steepness factors for all directions. The arrow indicates the
  ridge direction with the minimum steepness factor.}
\label{exam2}
\end{figure}
\subsection{Ridge flatness, $\beta$}

Smoothed optical map (middle panels of Fig.~\ref{exam1}) of a
relaxed cluster generally has a steep surface brightness profile in
all directions. Substructures appear in the smoothed map of an
unrelaxed cluster, and there usually exists a `ridge' extended from
the cluster center to a certain direction in the smoothed map.

We first get light profiles in various directions. For a given direct
$\phi$ ($0^{\circ}$ in the north, $90^{\circ}$ in the east), we fit
the light profile with a one-dimensional King model \citep{king62}
\begin{equation}
I_{\rm 1D}(r)=\frac{I_0}{1+(r/r_0)^2},
\end{equation}
where $r$ is the cluster centric distance in the given direction,
$I_0$ is the luminosity density at cluster center and $r_0$ is
characteristic radius of the one-dimensional Kind model. We now define
the steepness factor as $c_{200}=r_{200}/r_0$. For each cluster,
we calculate $c_{200}$ in 72 direction (5$^{\circ}$ each section) and
get the mean value of the steepness factor, $\langle c_{200}\rangle$.

The direction with the most flat profile is recognized as the ridge on
the map, which very likely indicates the axis of cluster merging and
has the minimum steepness factor, $c^R_{200}$. The ridge flatness is
defined as the relative steepness of the `ridge' to those of other
directions,
\begin{equation}
\beta=\frac{c^R_{200}}{\langle c_{200}\rangle}.
\end{equation}
For very relaxed clusters, the light profiles are similarly steep in
all directions, so that $\beta\sim 1$. For unrelaxed clusters, a
flatter ridge in the map gives a smaller $\beta$. For example, the
profiles of A2255 have similar steepness of $c_{200}\sim 12$ in most
directions (Fig.~\ref{exam2}), but in the ridge direction a much
small value of $c^R_{200}\sim 4$, and hence the cluster has a very
small of $\beta\sim 0.33$.

\subsection{Normalized deviation, $\delta$}

Relaxed clusters have very similarly smooth brightness profiles
in all directions, so that the smoothed optical map can be fitted by,
e.g., the two-dimensional elliptical King model. The unrelaxed
clusters have many substructures in the map, which result in more
deviation of the smoothed optical map from the model.

First, we model the smoothed map of a cluster within the region of
$r_{500}=2/3\;r_{200}$ with a two-dimensional elliptical King model of
\begin{equation}
I_{\rm 2D model}(x,y)=\frac{I_0}{1+(r_{\rm iso}/r_0)^2},
\end{equation}
where $I_0$ is the luminosity density at cluster center, $r_0$ is the
characteristic radius of the two-dimensional Kind model, $r_{\rm iso}$
is the cluster centric `distance' of an isophote,
\begin{equation}
r^2_{\rm iso}=[x\cos(\theta)+y\sin(\theta)]^2+\epsilon^2[-x\sin(\theta)+y\cos(\theta)]^2,
\end{equation}
where $\epsilon$ is the ratio of semiminor axis to semimajor axis of
an isophote ($\epsilon\le 1$), $\theta$ is the position angle of the
major axis. The normalized deviation $\delta$ of the residual map
after the model fitting is defined as
\begin{equation}
\delta=\frac{\sum_{i,j}[I(x_i,y_j)-I_{\rm 2D model}(x_i,y_j)]^2}{S^2},
\end{equation}
here $S$ is defined in Equation~\ref{eq_s}. For a relaxed cluster, the
smoothed map can be well fitted by a two-dimensional King model, so
that the deviation $\delta$ is small. A large $\delta$ means a larger
deviation from the best-fitting King model, which is an indication of
substructures produced by, e.g., violent mergers (see right panels of
Fig.~\ref{exam1}). 

\subsection{Uncertainties of $\alpha$, $\beta$ and $\delta$}

The uncertainties of $\alpha$, $\beta$ and $\delta$ of a galaxy
cluster mainly come from the contamination and incompleteness of
member galaxies. For the cluster sample in Table~\ref{tab1}, checking
with available spectroscopic data, we find that the contamination rate
and incompleteness for luminous member galaxies of $M^{\rm e}_r(z)\le-20.5$
within a radius of $r_{500}$ are about 10\%, 5\%, respectively.

We apply Monte Carlo simulations to estimate the uncertainties. 
For each cluster, a fraction of recognized member galaxies are
randomly selected and assumed to be contamination. After removing the
contamination, we get a new partially removed member galaxy data.
Considering the estimation of incompleteness on the uncertainties is
the same as that of contamination, we randomly remove a small fraction
(15\%) of the recognized member galaxies (not the BCG anyway). Follow
the calculations described above, we get values of $\alpha$, $\beta$
and $\delta$ for the partially removed dataset for 50 times. The
scatter of the values is taken as the uncertainty.

\begin{figure}
\centering
\includegraphics[width = 8.cm]{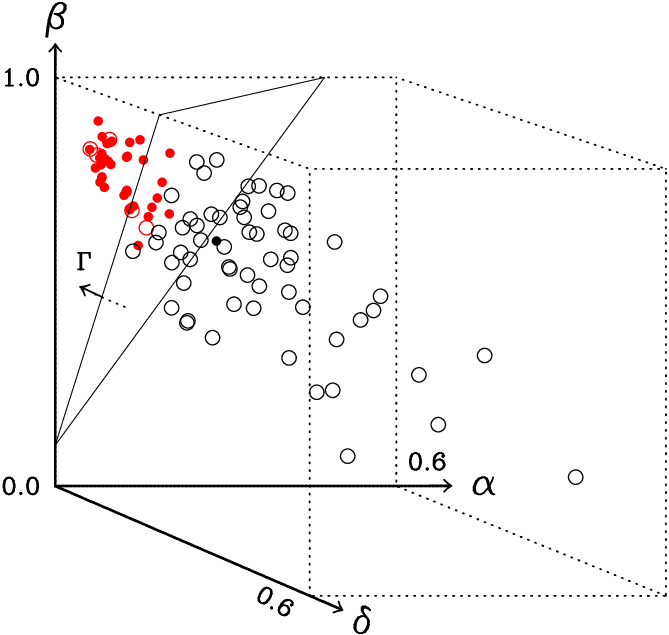}
\caption{In the three-dimensional space of $\alpha$, $\beta$ and
  $\delta$, relaxed clusters in Table~\ref{tab1} are shown by dots and
  unrelaxed clusters are shown by open circles. The $\alpha$, $\beta$
  and $\delta$ values are calculated with smooth scale parameters of
  $\sigma_a=0.03$, $\sigma_b=0.15$. The plane outlined by thin solid
  lines can separate the relaxed and unrelaxed clusters. Clusters
  farther than the plane are shown in red color. The relaxation
  parameter $\Gamma$ is defined as the distance to the plane.
}
\label{abc}
\end{figure}

\begin{figure}
\centering
\includegraphics[width = 8.cm]{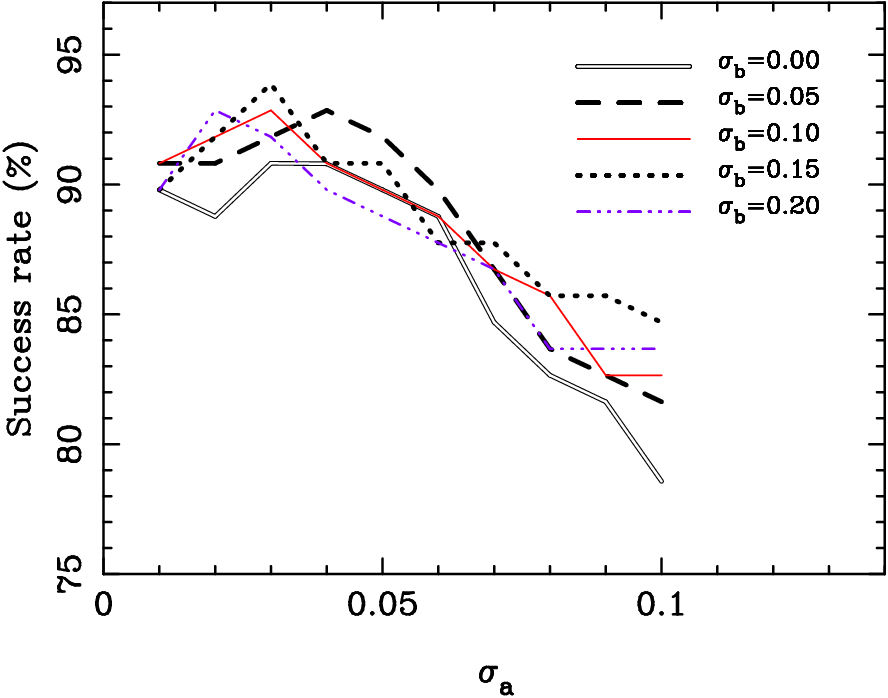}
\caption{Success rate for the separation of the relaxed and unrelaxed
  clusters in Table~\ref{tab1} as a function of $\sigma_a$. Different
  curves are plotted for different $\sigma_b$.  
}
\label{asyfit}
\end{figure}

\subsection{Relaxation parameter, $\Gamma$, and optimization}

Based on photometric data of a cluster, the $\alpha$, $\beta$ and
$\delta$ values can be calculated with properly assumed $\sigma_a$ and
$\sigma_b$. All these three parameters are related to the dynamical
state of a cluster, but they are not independent. We here seek the
best combination of the three parameters, and define the relaxation
parameter, $\Gamma$, for quantification of cluster dynamical state,
which should be able to effectively separate the relaxed and unrelaxed
states of the test clusters in Table~\ref{tab1}.

\begin{figure}
\centering
\includegraphics[width = 8.cm]{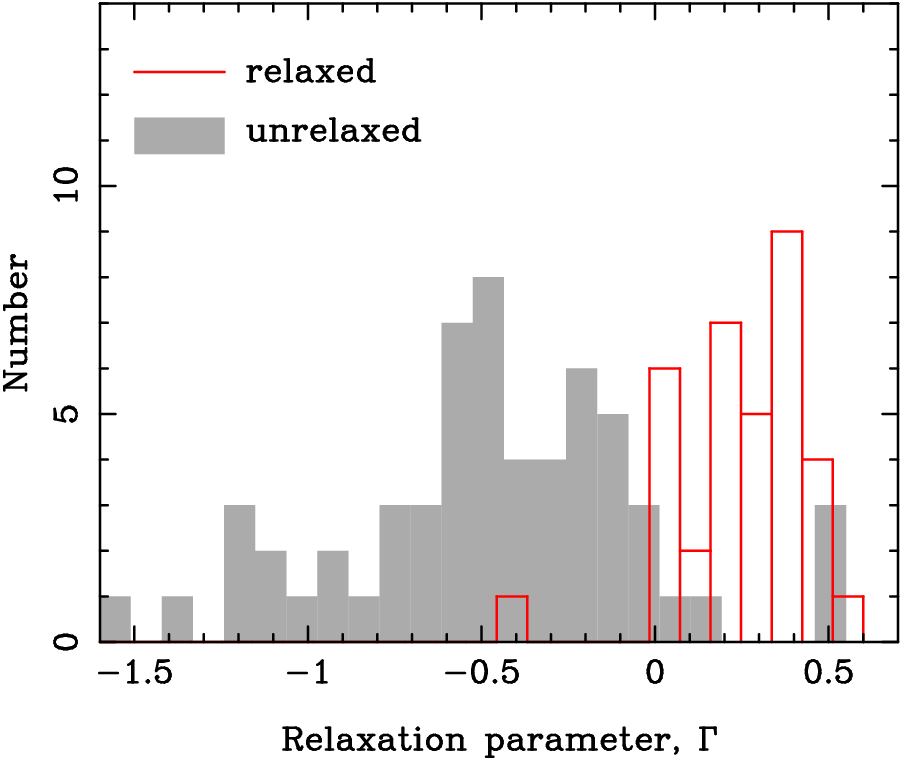}
\caption{Histogram for relaxation parameters of relaxed and unrelaxed
  clusters in Table~\ref{tab1}. 
}
\label{hist_gamma}
\end{figure}

First of all, for the smooth scale with a set of $\sigma_a$ and
$\sigma_b$, we obtain the smoothed maps of clusters in
Table~\ref{tab1}, and then calculate the quantities of $\alpha$,
$\beta$ and $\delta$. We find that a large smooth scale may erase the
merging features for unrelaxed clusters, while a small scale may
result in discrete patches in the smoothed map for misclassification.
Therefore, the optimal values of $\sigma_a$ and $\sigma_b$ should be
searched to separate the relaxed and unrelaxed clusters in the
three-dimensional space of $\alpha$, $\beta$ and $\delta$ (see
Fig.~\ref{abc}) by a plane which we will define below. 

Clearly, relaxed clusters in general have larger $\beta$ values and
smaller $\alpha$ and $\delta$ values than the unrelaxed clusters. The
data distribution in the three-dimensional space of $\alpha$, $\beta$
and $\delta$ in Fig.~\ref{abc} shows that the relaxed and unrelaxed
clusters can be separated by a plane:
\begin{equation}
\beta=A\,\alpha+B\,\delta+C. 
\label{sepa}
\end{equation}
The optimal plane should give the maximum rate of successful
separation. For any set of $\sigma_a$ and $\sigma_b$, we find the
optimal plane. After some iterations, we find that separation reaches
the maximum rate of 93.9\% for the relaxed and unrelaxed clusters in
Table~\ref{tab1} (see Fig.~\ref{asyfit}). The optimal value of
$\sigma_a$ is found to be 0.03 and that of $\sigma_b$ is 0.15, the
optimal plane has the parameters of $A=1.9$, $B=3.58$ and $C=0.1$.

Now, we define the relaxation parameter, $\Gamma$, to quantify
dynamical state of a cluster, which is the distance to the optimal
plane (see Fig.~\ref{abc}):
\begin{equation}
\Gamma =[\beta-(1.90\,\alpha+3.58\,\delta+0.10)]/k,
\label{gamma}
\end{equation}
here $k=\sqrt{1+A^2+B^2}$. In practical calculation, we ignore the
constant factor $k$ (i.e., $k=1$). The relaxation parameter is
positive for relaxed clusters and negative for unrelaxed clusters. A
larger $\Gamma$ means that a cluster appears more relaxed.

\begin{figure*}
\centering
\resizebox{55mm}{!}{\includegraphics{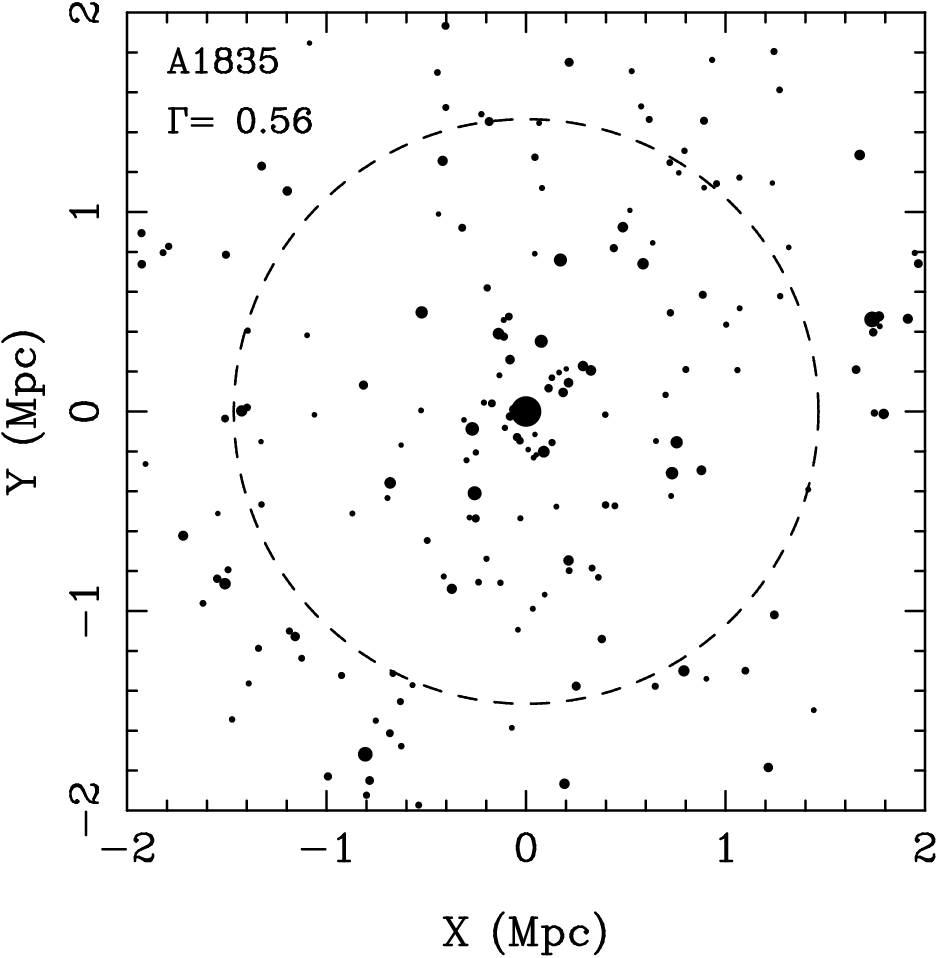}}~%
\resizebox{55mm}{!}{\includegraphics{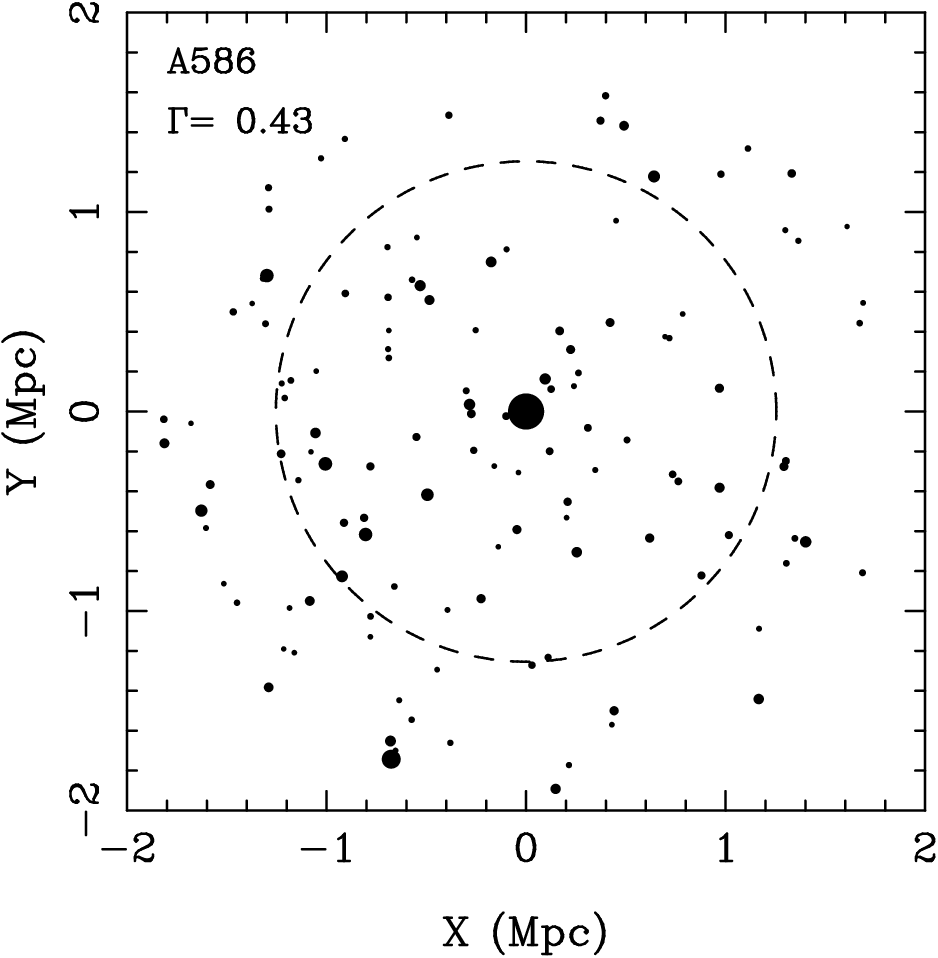}}~%
\resizebox{55mm}{!}{\includegraphics{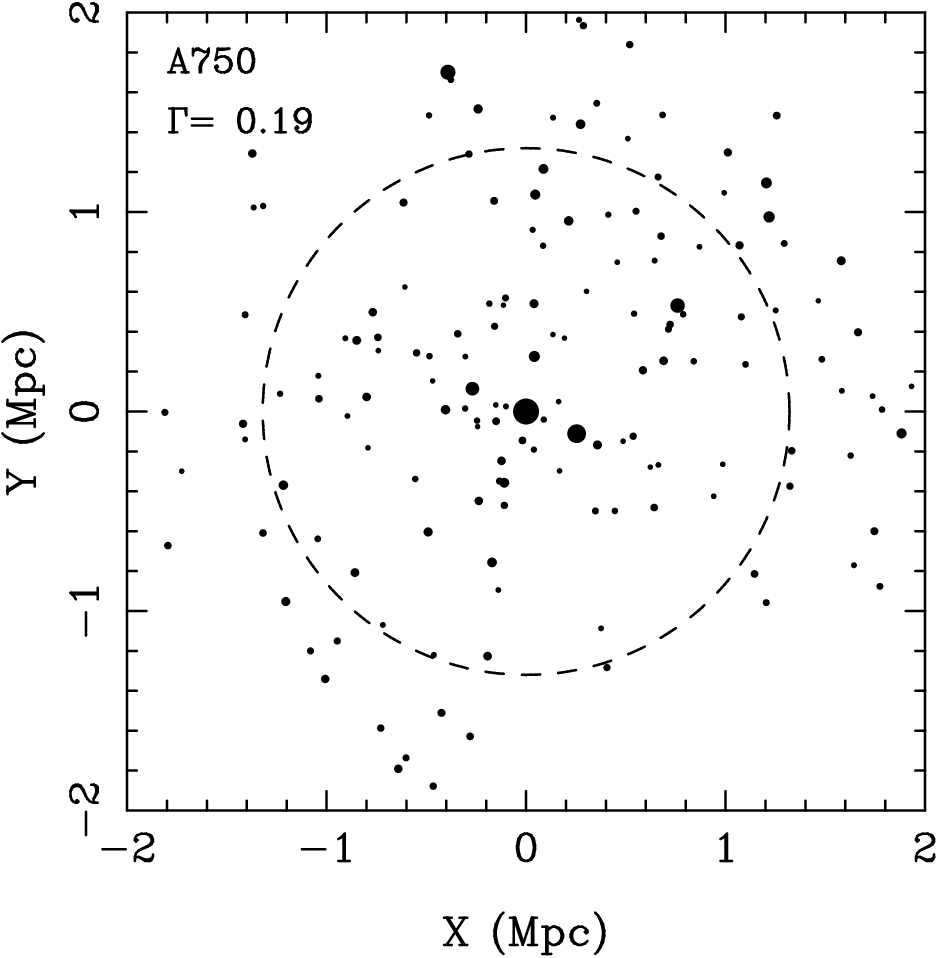}}\\[0.5mm]
\resizebox{55mm}{!}{\includegraphics{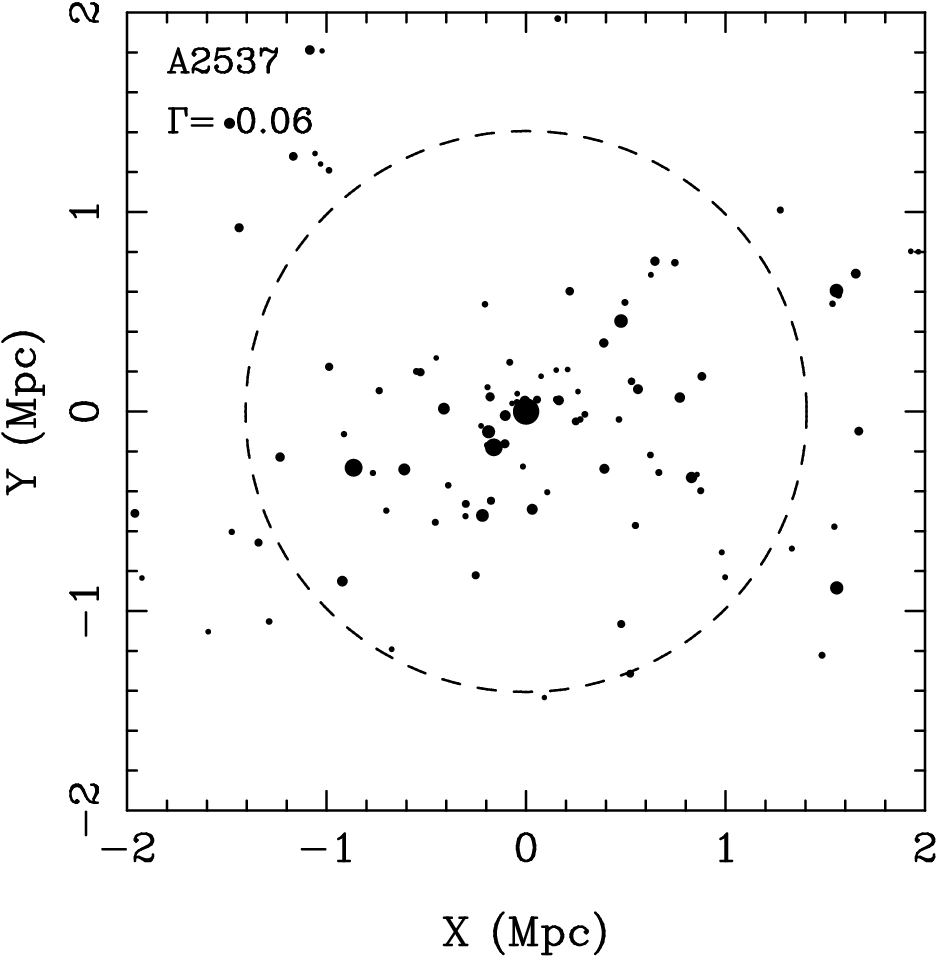}}~%
\resizebox{55mm}{!}{\includegraphics{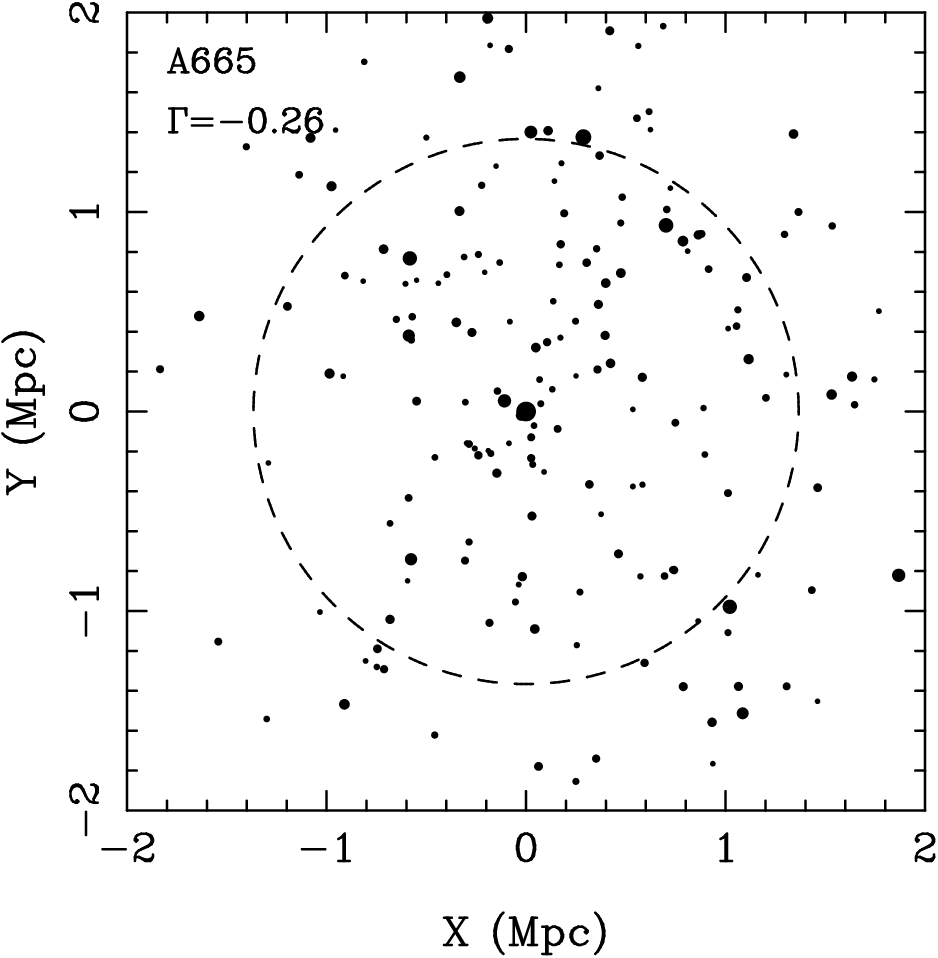}}~%
\resizebox{55mm}{!}{\includegraphics{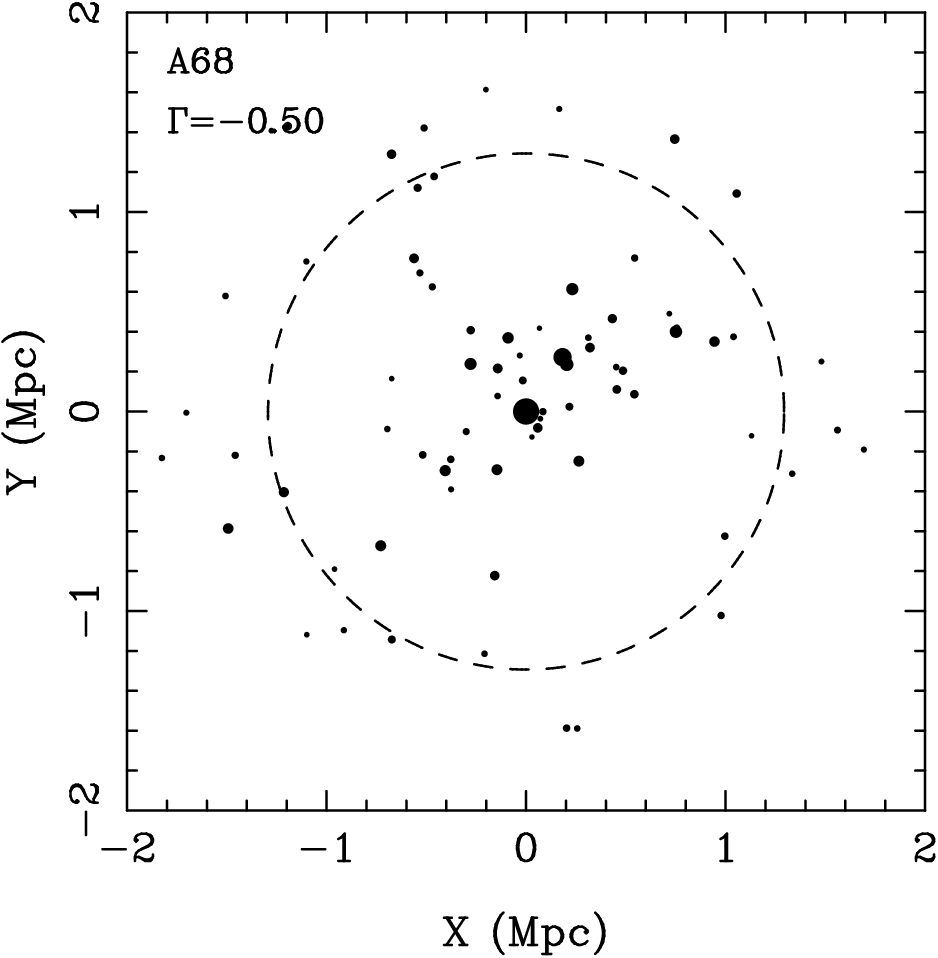}}\\[0.5mm]
\resizebox{55mm}{!}{\includegraphics{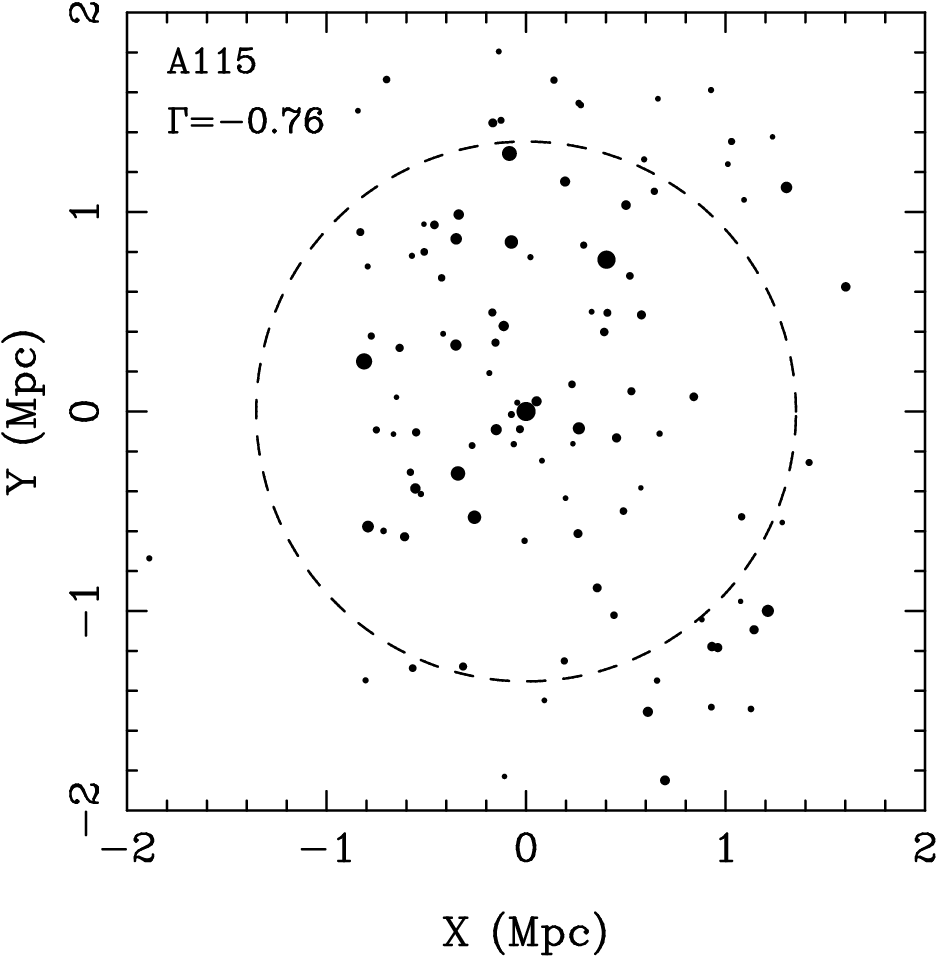}}~%
\resizebox{55mm}{!}{\includegraphics{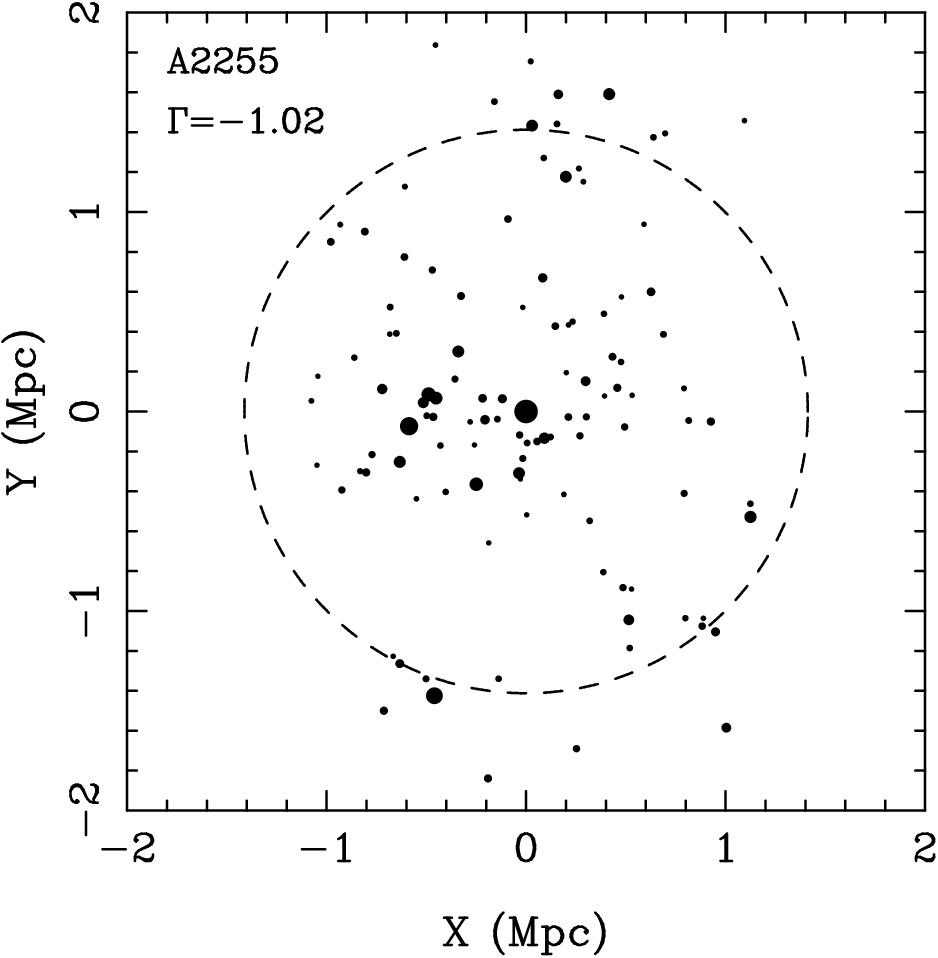}}~%
\resizebox{55mm}{!}{\includegraphics{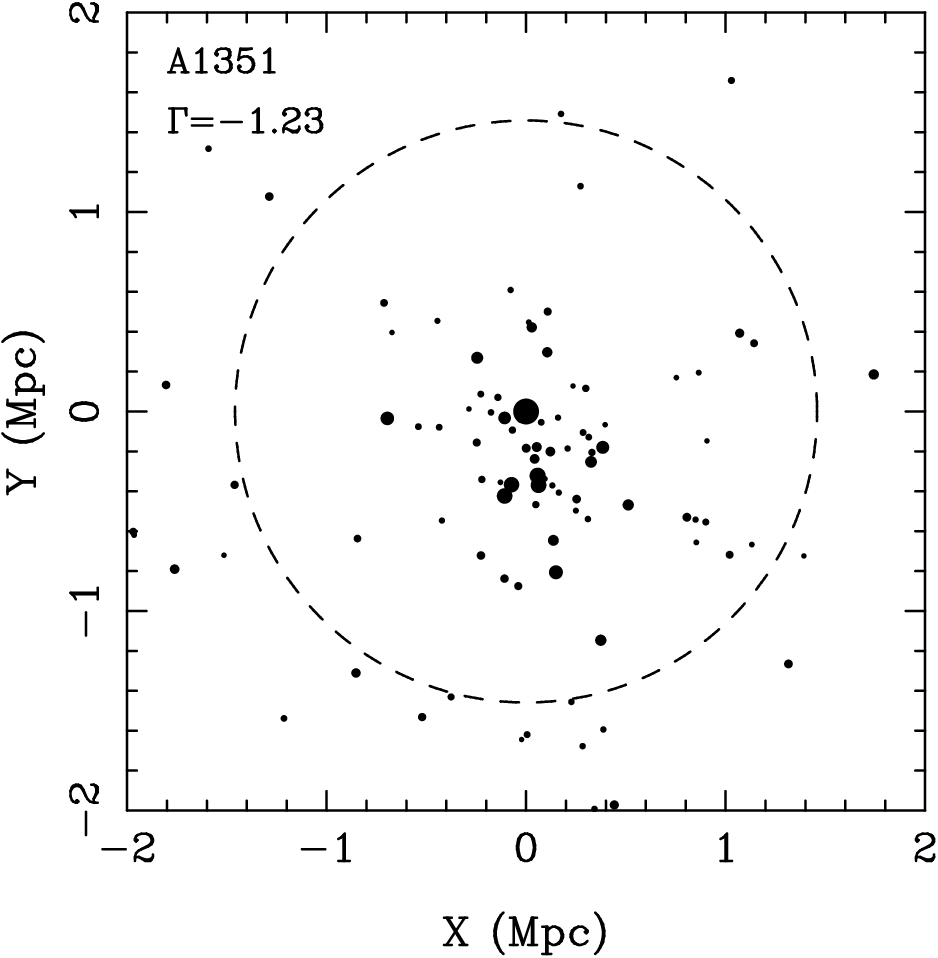}}\\[0.5mm]
\resizebox{55mm}{!}{\includegraphics{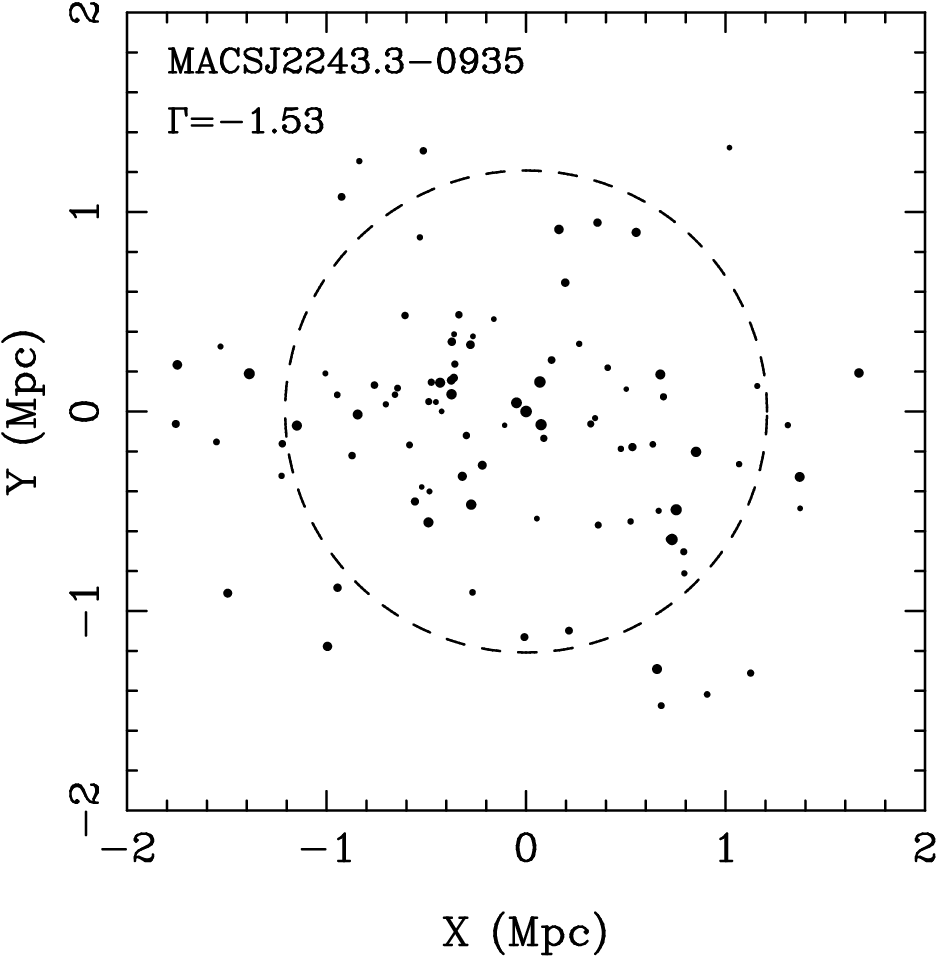}}~%
\resizebox{55mm}{!}{\includegraphics{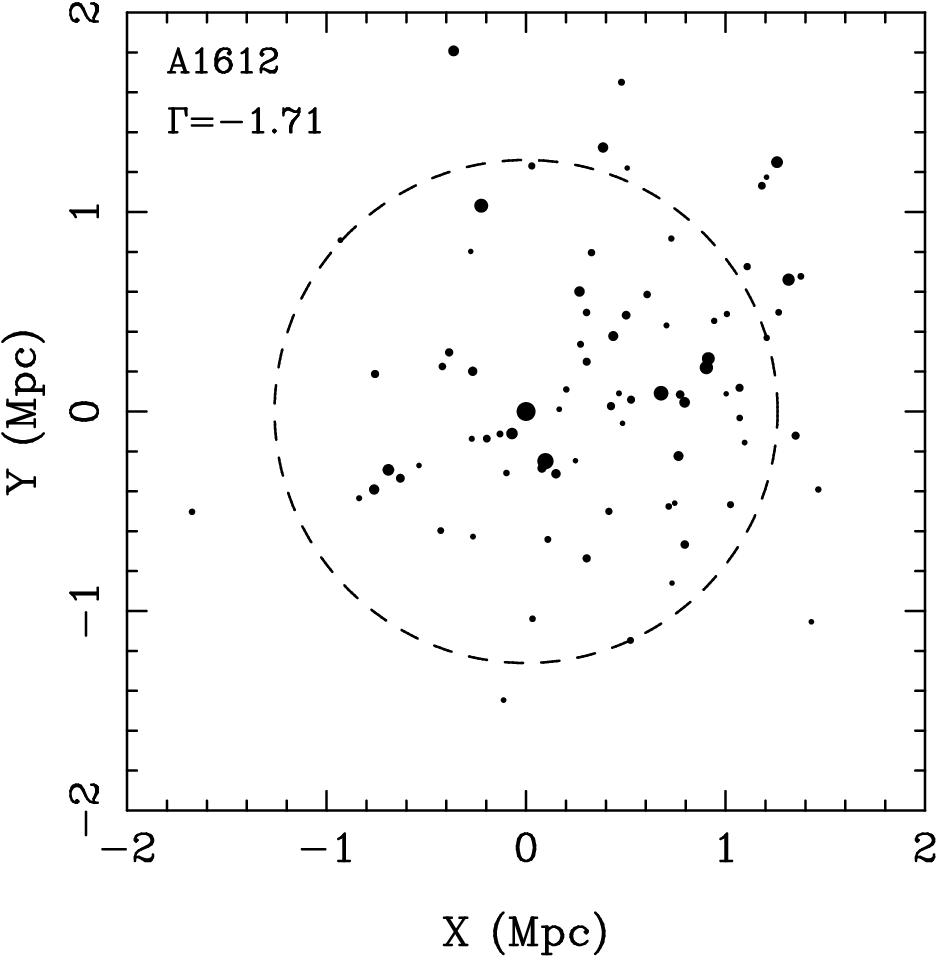}}~%
\resizebox{55mm}{!}{\includegraphics{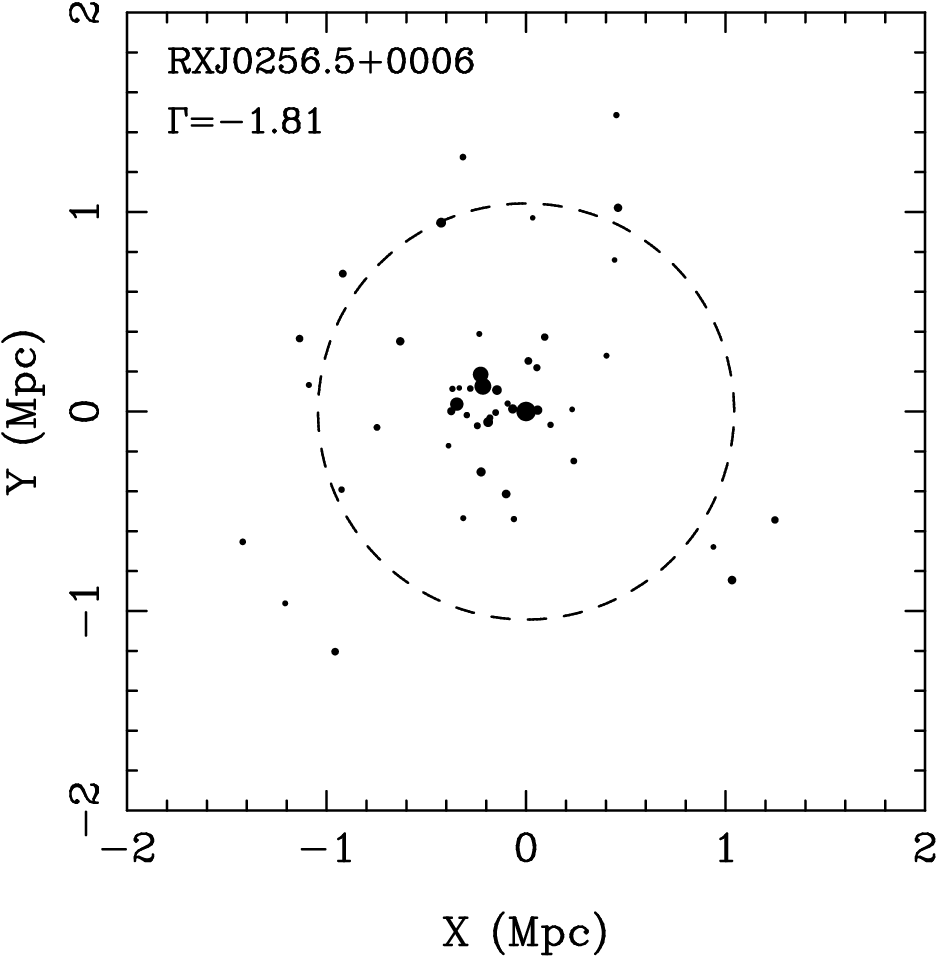}}
\caption{Distribution of cluster member galaxies of 12 example
  clusters with different $\Gamma$ values in the range of $-2\lesssim
  \Gamma<0.6$.  Sizes of black dots are scaled by the square root of
  galaxy luminosities. A large positive $\Gamma$ indicates the relaxed
  state of clusters, and a very negative $\Gamma$ indicate the very
  unrelaxed state. Large circles indicate the central region of
  clusters with a radius of $r_{500}$.
\label{exam3}}
\end{figure*}

We have calculated the values for $\alpha$, $\beta$, $\delta$ and
$\Gamma$ for clusters in Table~\ref{tab1}. The relaxed and
unrelaxed clusters are well separated in the three-dimensional space
by the plane shown in Fig.~\ref{abc}, see also the histogram in
Fig.~\ref{hist_gamma}. A few exceptions are discussed in
Appendix. As shown by examples in Fig.~\ref{exam3}, clusters with a
low negative value of $\Gamma$ have irregular distributions of member
galaxies and no dominant central galaxies.

In some merging clusters, BCGs are not located in the cluster center
\citep{whl09,evn+12}. Our calculations for the center on the BCGs give
very low value of $\beta$ and high values of $\alpha$ and $\delta$,
and then such clusters are recognized as unrelaxed with a very
negative $\Gamma$.

\begin{figure}
\centering
\includegraphics[width = 7.5cm]{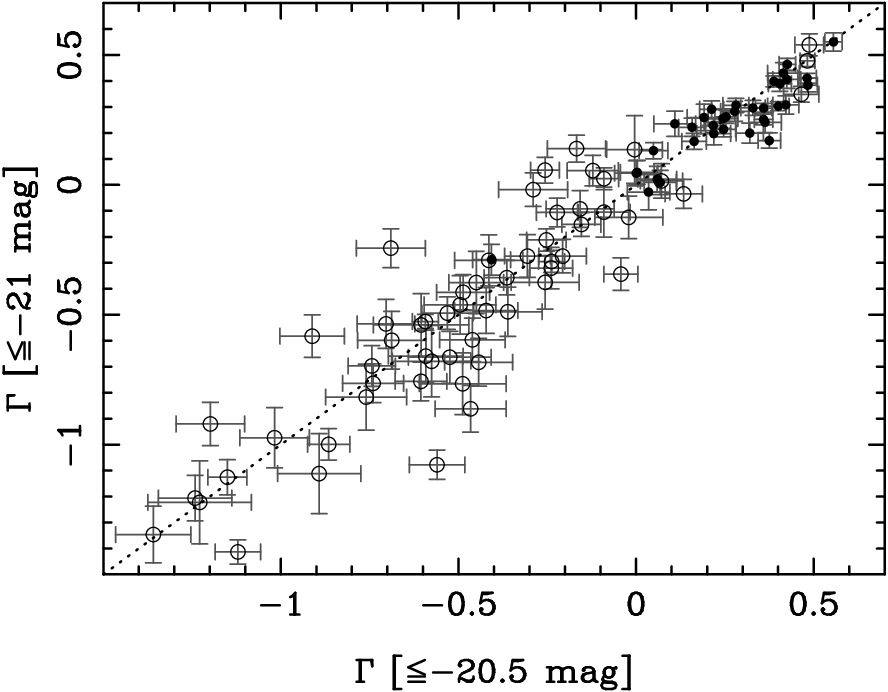}
\includegraphics[width = 7.5cm]{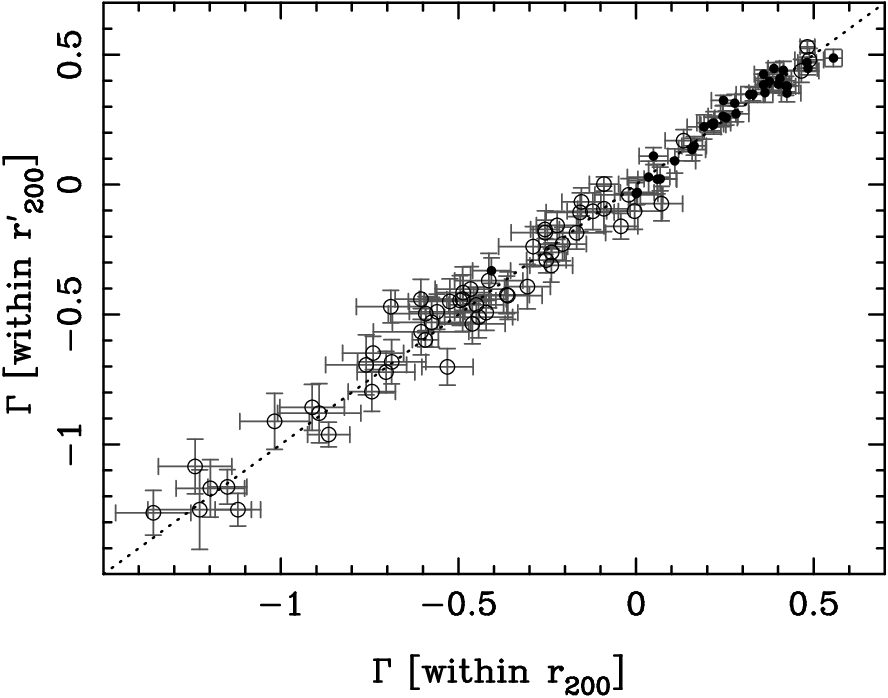}
\caption{Comparison of relaxation parameters of clusters in
  Table~\ref{tab1} calculated with different thresholds for absolute
  magnitudes of member galaxies (upper panel) or different estimates
  for cluster radii (lower panel). Relaxed clusters are plotted as
  black dots and unrelaxed clusters as open circles.}
\label{mag}
\end{figure}

\subsection{Tolerance of the relaxation parameter on faint member galaxies 
and cluster radius}

In definition of the relaxation parameter of a cluster, we have
taken the luminosity threshold of member galaxies as an absolute
magnitude of $M^{\rm e}_r=-20.5$. Now we investigate the tolerance of the
relaxation parameter by discarding the faint member galaxies and
setting the threshold $M^{\rm e}_r=-21.0$, so that $\alpha$, $\delta$,
$\beta$ and $\Gamma$ values are calculated by using less member
galaxies. For clusters in Table~\ref{tab1}, we find that the
$\Gamma$ values are very consistent with each other for the two
thresholds (the upper panel of Fig.~\ref{mag}), which suggests that
the incompleteness of faint member galaxies does not affect the
estimation of relaxation parameter, $\Gamma$, as long as there are
enough luminous member galaxies used for calculations.

The relaxation parameter, $\Gamma$, is derived from $\alpha$, $\delta$
and $\beta$ within the central region of an area of $r_{500}=2/3
r_{200}$. It is also possible that $r_{200}$ has some uncertainty. We
investigate the tolerance of relaxation parameters on the cluster
radius, by assuming that $r_{200}$ in Table~\ref{tab1} is
systematically overestimated so that the real radius
$r'_{200}=0.9r_{200}$. With the assumed smaller cluster radii, some
member galaxies in the outer region are discarded and the smooth
scales also become systematically smaller than the original values
according to Equation~(\ref{sscale}). We calculate $\alpha$, $\delta$
and $\beta$ and $\Gamma$, and find very good agreement between
original and new $\Gamma$ values (the lower panel of
Fig.~\ref{mag}).

\subsection{Comparison of $\Gamma$ with dynamical state parameters 
estimated from X-ray data}
\label{dynxray}

Previously, substructures and dynamical states of clusters were often
estimated from X-ray data, as represented by the concentration
\citep[e.g.][]{srt+08}, the centroid shift \citep[e.g.][]{mef+95} and
the power ratio \citep[e.g.][]{bt95} derived from X-ray image. Cooling
time of hot gas derived from X-ray spectra sometimes was also used as
indication of dynamical states \citep[e.g.][]{vf04,bfs+05}. Now we
compare the relaxation parameters with the dynamical parameters
derived from X-ray data.

\begin{figure}
\centering
\includegraphics[width = 8.cm]{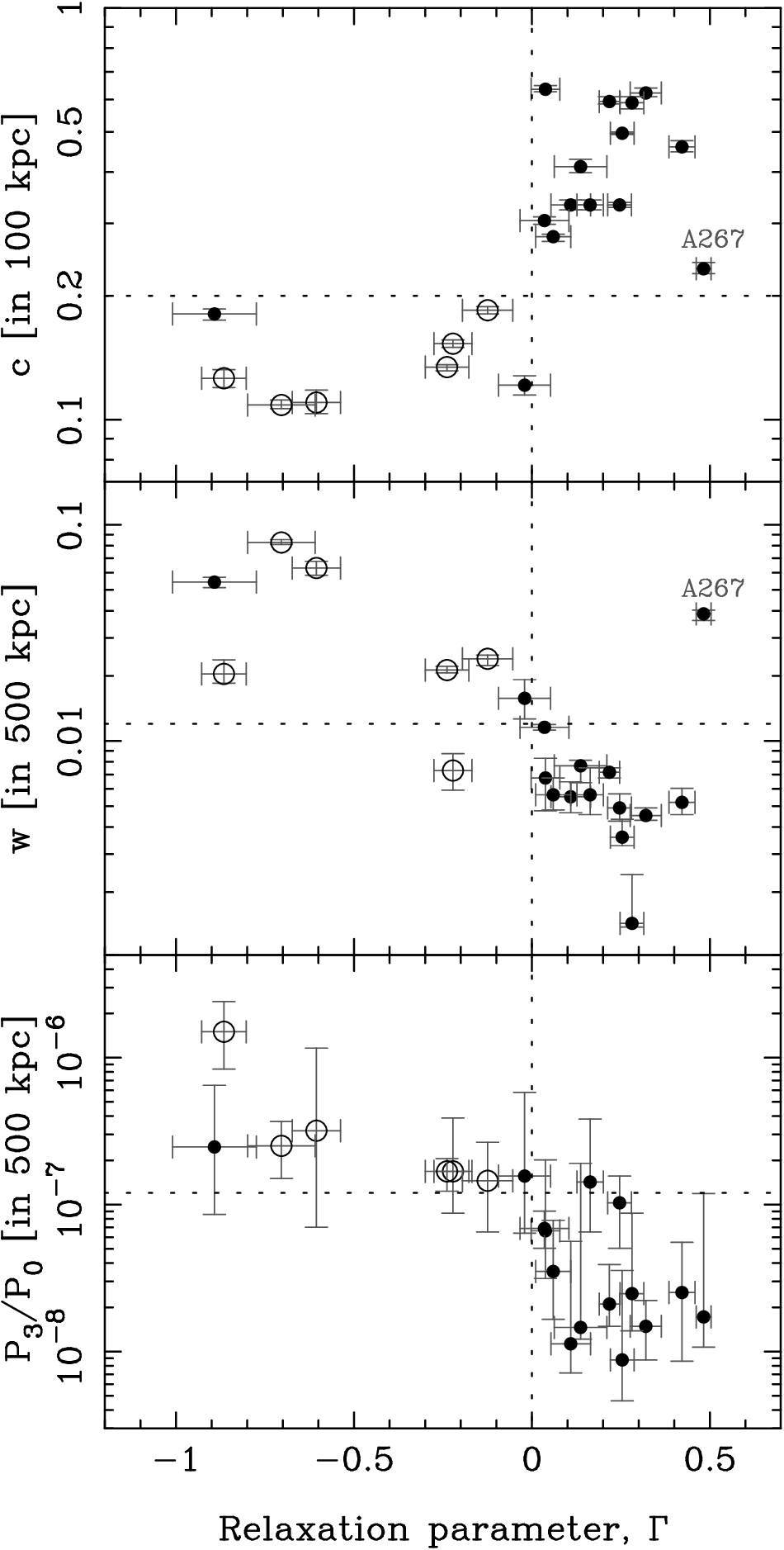}
\caption{Correlation between the relaxation parameter, $\Gamma$, with
  the concentration $c$ (upper panel), centroid shift $w$ (middle
  panel) and the power ratio $P_3/P_0$ (lower panel) derived from
  X-ray data by \citet{ceg+10} for 21 clusters. The black dots are
  clusters without radio halo detections, and open circles are
  clusters with radio halo detections. The horizontal dotted lines
  indicate the separation between radio halo and no-radio halo
  clusters by \citet{ceg+10}, the vertical line indicates the
  separation between relaxed and unrelaxed clusters by $\Gamma$.}
\label{comp1}
\end{figure}

The first dataset of X-ray dynamical parameters are taken from
\citet{ceg+10} who have the concentration, centroid shift and power
ratio measurements of X-ray images published for 32 clusters.  There
are 21 clusters in the SDSS-III region (A267, A611, A697, A773, A781,
A1423, A1682, A1758, A2219, A2261, A2390, A2537, A2631, MACS
J1115.8+0129, MACS J2228.5+2036, RXC J0437.1+0043, RXC J1504.1$-$0248,
RX J1532.9+3021, Z2089, Z2701, Z7160), 19 of which have been listed in
Table~\ref{tab1}. We also calculate the relaxation parameters for the
other two clusters (RXCJ0437.1+0043, Z2089). In this sample of 21
clusters, six of them have radio halos detected and are most probably
merging clusters, and other 15 have not. Comparisons of the relaxation
parameter, $\Gamma$, with the concentration $c$, centroid shift $w$
and power ratio $P_3/P_0$ are shown in Fig.~\ref{comp1}. The
correlations are reasonable except for A267. The clusters with and
without radio halos can be well separated.

\begin{figure}
\centering
\includegraphics[width = 8.cm]{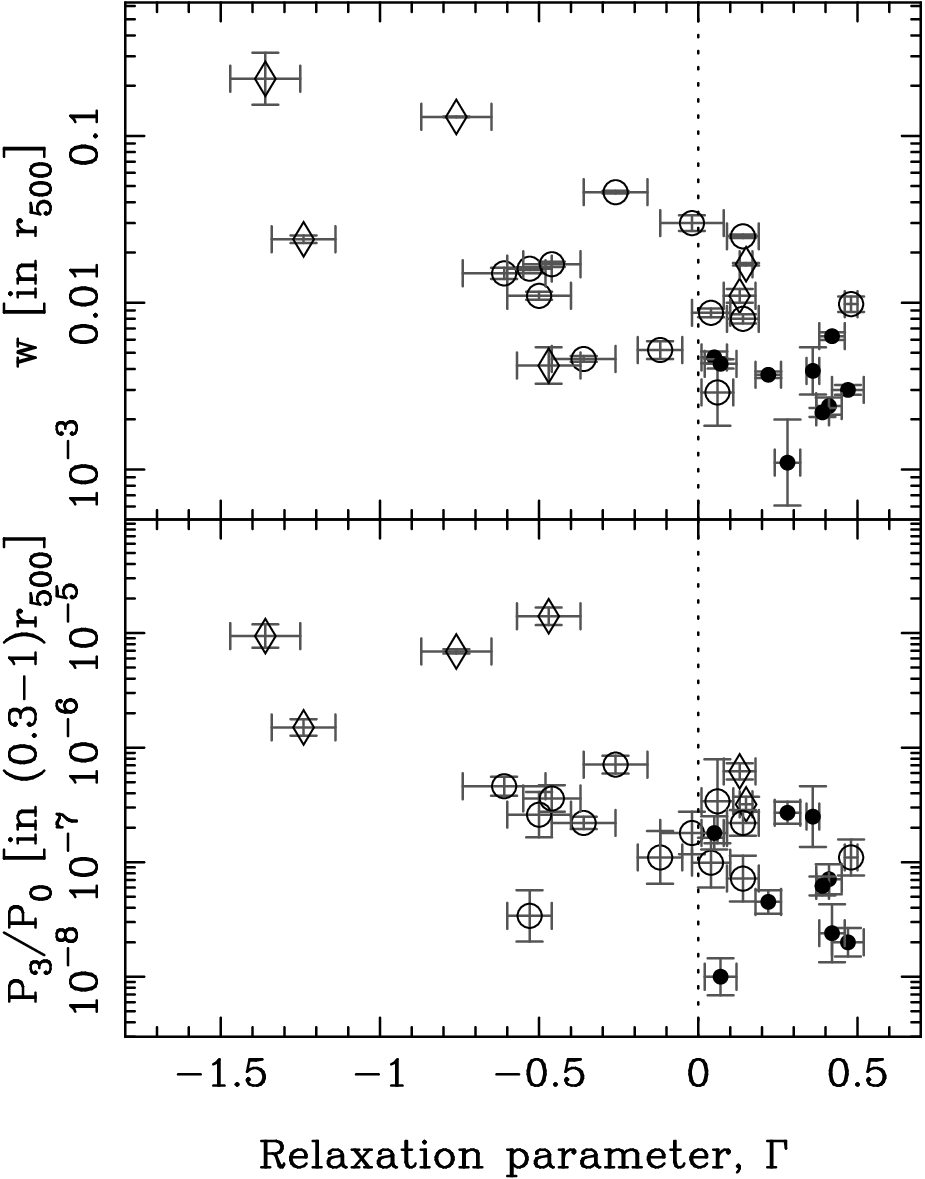}
\caption{Similar to Fig.~\ref{comp1} but for the dynamical
  parameters ($w$ and $P_3/P_0$) of 28 X-ray clusters derived by
  \citet{wbs+13}, who have visually classified the clusters into
  `regular' (black dots), `intermediate' (open circle), `double'
  and `complex' (diamond) morphology categories.}
\label{comp2}
\end{figure}

The second dataset of X-ray dynamical parameters are taken from
\citet{wbs+13}. In their sample, there are 28 rich clusters in the
area of the SDSS-III with an optical richness of $R_{L\ast}\ge50$
(A13, A68, A115, A267, A383, A665, A773, A963, A1413, A1589, A1689,
A1763, A1775, A1914, A2065, A2626, A2390, A2537, A2631, RXC
J0003.8+0203, RXC J0821.8+0112, RXC J1302.8$-$0230, RXC J1516.3+0005,
RXC J1516.5$-$0056, RXC J2157.4$-$0747, RXC J2129.6+0006,Z3146,
Z7160), among which 25 clusters are in Table~\ref{tab1}. We get the
relaxation parameters $\Gamma$ for other three clusters (A1775,
RXCJ1302.8-0230, RXCJ1516.3+0005). Reasonable correlations between our
relaxation parameters with dynamical parameters ($w$ and $P_3/P_0$)
are shown in Fig.~\ref{comp2}.
 
\begin{figure}
\centering
\includegraphics[width = 8.cm]{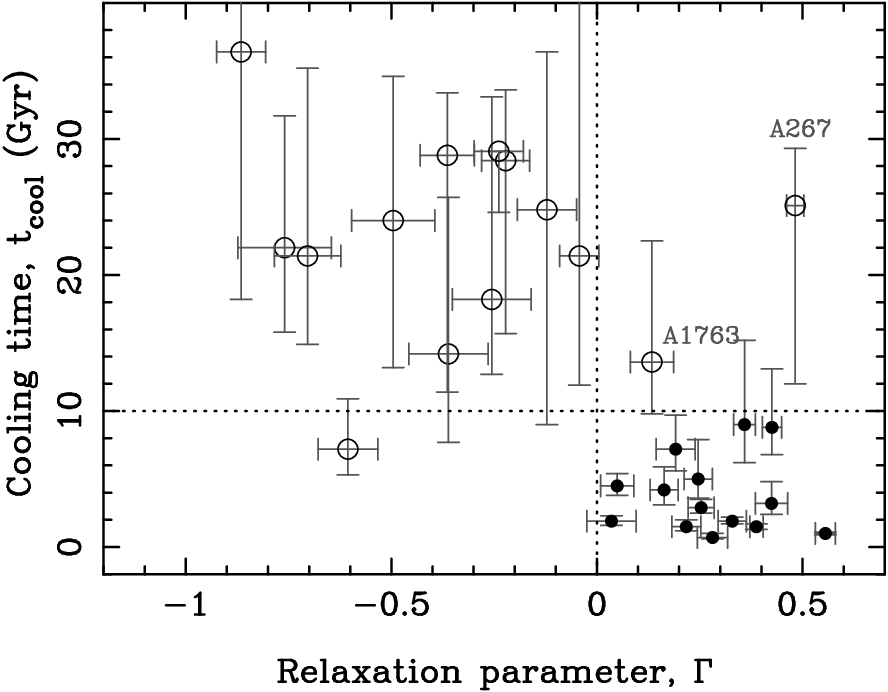}
\caption{Comparison between the relaxation parameter, $\Gamma$, and
  cooling time at a radius of 50 kpc for 28 clusters derived from
  X-ray data \citep{bfs+05}. The black dots and open circles indicate
  relaxed and unrelaxed clusters, respectively. The horizontal and
  vertical dotted lines indicate $\Gamma=0$ and $t_{\rm cool}=10$ Gyr.}
\label{cool}
\end{figure}

Previous studies show that relaxed clusters have shorter cooling time
than unrelaxed clusters \citep[e.g.][]{fab94,vf04,bfs+05,pf06}. The
cooling times of 28 clusters in the SDSS-III region (A68, A115, A267,
A586, A665, A697, A750, A773, A781, A963, A1423, A1682, A1758, A1763,
A1835, A1914, A2111, A2219, A2259, A2261, A2390, RX J1532.9+3021, RX
J1720.1+2638, RX J2129.6+0006, Z1953, Z2701, Z3146, Z7160) have been
derived from X-ray data by \citet{bfs+05}. All these clusters are
included in Table~\ref{tab1}. Fig.~\ref{cool} shows that the
relaxation parameter $\Gamma$ and cooling time $t_{\rm cool}$ at a
radius of 50 kpc are well correlated. Almost all clusters with cool
cores of $t_{\rm cool}\le10$~Gyr have $\Gamma>0$.

\begin{figure}
\centering
\includegraphics[width = 8.cm]{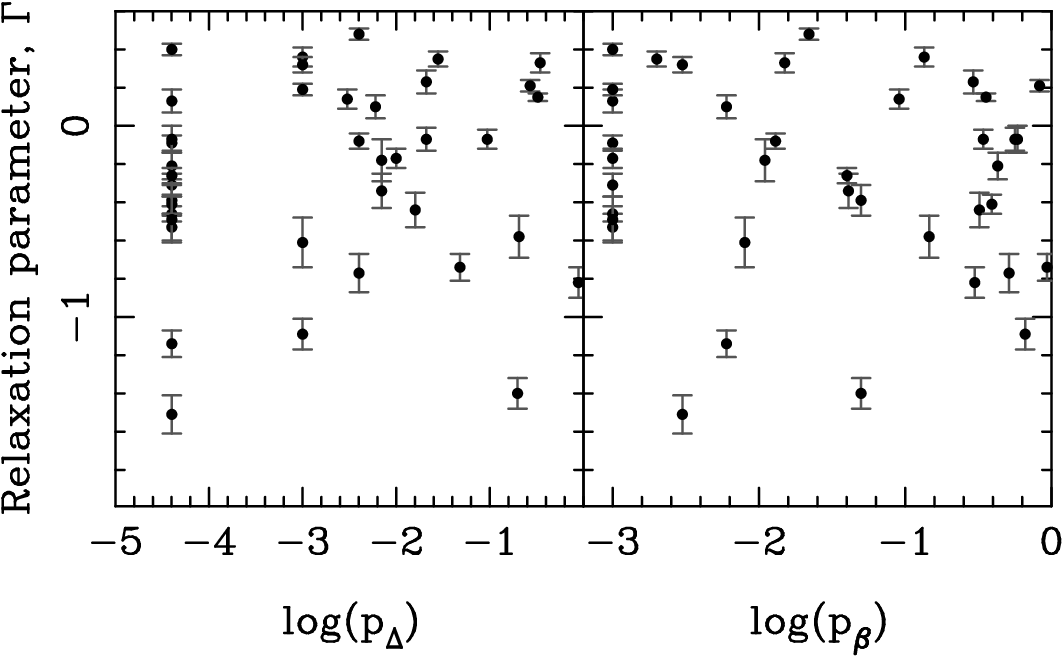}
\caption{Comparison between the relaxation parameter, $\Gamma$, with
  the substructure significances by three-dimensional
  Dressler-Shectman test (left panel) and two-dimensional $\beta$
  (i.e., asymmetry) test derived by \citet{evn+12}.}
\label{comp3}
\end{figure}

\subsection{Comparison of $\Gamma$ with previous optical study of 
substructures}

Based on the SDSS spectroscopic data, \citet{evn+12} searched for
substructures in 109 rich clusters of \citet{ttl12} by using a number
of one-, two- and three-dimensional tests. Their three-dimensional tests
include the Dressler--Shectman test \citep{ds88} and the $\alpha$ test
which measures the centroid shift of cluster galaxies weighted by
local velocity dispersion \citep{wb90}. Their two-dimensional test
includes the $\beta$ test which measures the asymmetry of galaxy
distribution \citep{wod88}. These tests give $p$-values to quantify
substructure significances. Smaller $p$-value means larger probability
of substructure.
Among the \citet{evn+12} sample, 13 clusters (A1589, A1650, A1750,
A1795, A1904, A1991, A2029, A2048, A2061, A2065, A2069, A2142, A2244)
are included in our Table~\ref{tab1} and 26 clusters (A628, A671,
A933, A1066, A1205, A1307, A1516, A1541, A1552, A1569, A1663, A1691,
A1767, A1775, A1809, A1831, A1913, A1999, A2020, A2022, A2079, A2124,
A2175, A2245, A2249, J141735.5+020312) in our Table~\ref{tab2}
(Section~\ref{dis_gamma}).

\begin{table*}
\begin{minipage}{160mm}
\caption[]{Relaxation parameters of 2092 rich clusters in the SDSS 
(see online Supporting Information for the full table).}
\begin{center}
\begin{tabular}{rrrcccrrrrc}
\hline
\mc{1}{c}{Name}    &\mc{1}{c}{R.A.}       &\mc{1}{c}{Dec.}   &\mc{1}{c}{$z$}   & flag$_z$  &\mc{1}{c}{$r_{\rm BCG}$} &
\mc{1}{c}{$r_{200}$}&\mc{1}{c}{$R_{L\ast}$}&\mc{1}{c}{$N_{200}$}&\mc{1}{c}{$\Gamma$}&\mc{1}{c}{$\sigma_{\Gamma}$} \\
                   &\mc{1}{c}{(deg)}      &\mc{1}{c}{(deg)}   &                   &                        & &
\mc{1}{c}{(Mpc)}   &                      &                   &                   &                         \\
\mc{1}{c}{(1)}     &\mc{1}{c}{(2)}        &\mc{1}{c}{(3)}     &\mc{1}{c}{(4)}     &\mc{1}{c}{(5)}          &
\mc{1}{c}{(6)}     &\mc{1}{c}{(7)}        &\mc{1}{c}{(8)}     &\mc{1}{c}{(9)}     &\mc{1}{c}{(10)}         &\mc{1}{c}{(11)}\\
\hline
WHL J000012.6$+$103806&    0.05238&   10.63509&  0.1824&  0&  16.70&  1.43&  56.77&  38&  $-1.67$&   0.10\\ 
WHL J000021.7$+$150612&    0.09053&   15.10328&  0.2991&  0&  17.72&  1.47&  54.05&  38&  $-0.09$&   0.06\\
WHL J000026.3$+$215405&    0.10958&   21.90143&  0.1665&  0&  16.17&  1.52&  57.68&  39&  $-0.30$&   0.08\\ 
WHL J000039.9$+$300305&    0.16637&   30.05137&  0.1906&  0&  16.51&  1.48&  54.85&  39&  $-0.15$&   0.06\\ 
WHL J000111.5$+$213213&    0.29792&   21.53696&  0.4000&  0&  18.75&  1.53&  68.03&  55&  $-1.06$&   0.11\\ 
WHL J000117.2$-$031648&    0.32183& $-3.28003$&  0.2974&  0&  17.74&  1.43&  57.24&  37&  $-0.72$&   0.14\\ 
WHL J000126.3$-$000143&    0.35969& $-0.02867$&  0.2465&  1&  16.93&  1.46&  58.61&  34&  $-0.33$&   0.06\\ 
WHL J000158.5$+$120358&    0.49367&   12.06612&  0.2086&  0&  15.91&  1.75&  88.28&  65&  $ 0.55$&   0.04\\ 
WHL J000311.6$-$060530&    0.79826& $-6.09169$&  0.2484&  0&  16.72&  1.79& 115.74&  88&  $-0.69$&   0.06\\ 
WHL J000318.1$+$043739&    0.82531&    4.62755&  0.0989&  0&  14.18&  1.37&  64.51&  32&  $ 0.52$&   0.03\\
\hline
\end{tabular}
\end{center}
{Note.  
Column (1): cluster name with J2000 coordinates of cluster. 
Column (2): R.A. (J2000) of BCG.
Column (3): Dec. (J2000) of BCG. 
Column (4): redshift. It is spectroscopic redshift if the flag 
in Column (5) is `1', or photometric redshift if the flag is `0'. 
Column (5):flag of redshift.
Column (6): $r$-band magnitude of BCG.
Column (7): $r_{200}$ of cluster (Mpc). 
Column (8): cluster richness.
Column (9): number of member galaxies within $r_{200}$.
Column (10) and (11): relaxation parameter and uncertainty.
}
\label{tab2}
\end{minipage}
\end{table*}

Fig.~\ref{comp3} shows the comparison between the relaxation
parameter, $\Gamma$, with the substructure significances estimated by
three-dimensional Dressler-Shectman test and two-dimensional $\beta$
test for the 39 clusters in \citet{evn+12}. We do not find significant
correlations, which is somehow unexpected. Our calculations are based
on the smoothed map of member galaxy distribution weighting their
luminosities. The relaxation parameter we derive reflects the
substructure of brightness distribution of member galaxies within the
radius $r_{500}$, so that it has good correlations with dynamical
state parameters from X-ray data (see Section~\ref{dynxray}). We
notice that the cluster sizes \citet{evn+12} work on are a few
($\sim$5--7) times of cluster virial radii for above 39 clusters
\citep{ttl12}. The substructure significances they estimated may
reflect the global substructures for position and velocity
distributions of member galaxies within a much larger region than
$r_{500}$.

\section{Dynamical states of 2092 rich clusters}

Compared to the dynamical parameters derived from X-ray data for
clusters, the relaxation parameter we defined has the advantage that
it can be easily estimated from optical photometric data for positions
and optical magnitude of member galaxies. It has successfully
separated the known relaxed and unrelaxed clusters of the test sample
of rich clusters with a rate of 94\%. We can apply the method to all
rich clusters to diagnose their dynamical state of clusters
whenever the optical photometric data for member galaxies are
available.

\citet{whl12} have identified 132\,684 clusters in the redshift range
of $0.05 < z < 0.8$ from the SDSS-III. Using photometric redshifts of
galaxies, we recognized a cluster when the richness $R_{L\ast}\ge12$
and the number of member galaxies $N_{200}\ge8$ within a photometric
redshift slice of $z \pm 0.04(1 + z)$ and a radius of $r_{200}$. In
this work, the spectroscopic redshifts of clusters are adopted if
available, otherwise photometric redshifts are used. We have used the
spectroscopic data of the SDSS DR9 \citep{dr9+12} to update member
galaxy list, $r_{200}$ and richness estimates for galaxy clusters
taken from the catalog of \citet{whl12}. Galaxies are removed from the
member galaxy list if they have a velocity difference $\Delta v>2500$
km~s$^{-1}$ from cluster redshift and the missing galaxies in the
photo-$z$ data are included if they have a velocity difference $\Delta
v\le2500$ km~s$^{-1}$ from cluster redshift. The cluster richness,
$R_{L\ast}$, and radius, $r_{200}$, are then recalculated.

In this section, we quantify the dynamical states for 2092 rich
clusters (Table~\ref{tab2} ) in the redshift range of $0.05<z\le0.42$
with a richness of $R_{L\ast}\ge50$. The redshift range is selected to
make the cluster sample and the member galaxies approximately
volume-limited complete \citep{whl12}. Above the richness, the
reliability of cluster identification is nearly 100\%.

\subsection{Distribution of relaxation parameters, $\Gamma$}
\label{dis_gamma}

\begin{figure}
\centering
\includegraphics[width = 7.5cm]{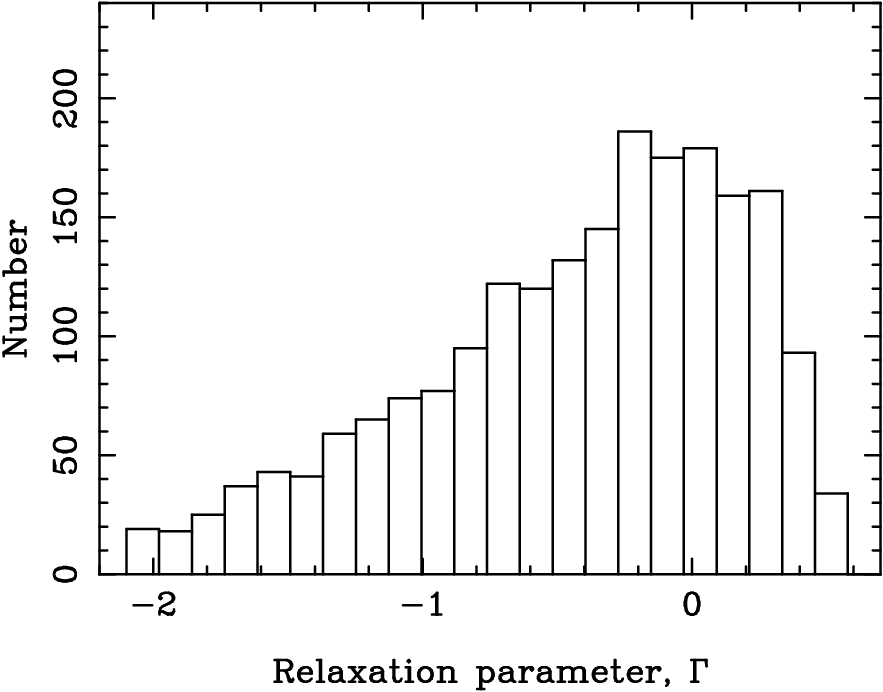}
\caption{Histogram distribution of relaxation parameters for 2092 rich
  clusters.}
\label{histrex}
\end{figure}

The relaxation parameters $\Gamma$ of these 2092 rich clusters are
calculated from $\alpha$, $\beta$ and $\delta$ by using the smoothed
photometric data of the SDSS. Fig.~\ref{histrex} shows the histogram
distribution of relaxation parameters for 2092 rich clusters. The
values of $\Gamma$ have a continuous distribution in the range of
$-2.0\lesssim\Gamma<0.6$ with a peak at $\Gamma\sim0$, which indicates
that most clusters have intermediate dynamical states, rather than
clearly very relaxed or very unrelaxed.

\begin{figure}
\centering
\includegraphics[width = 7.5cm]{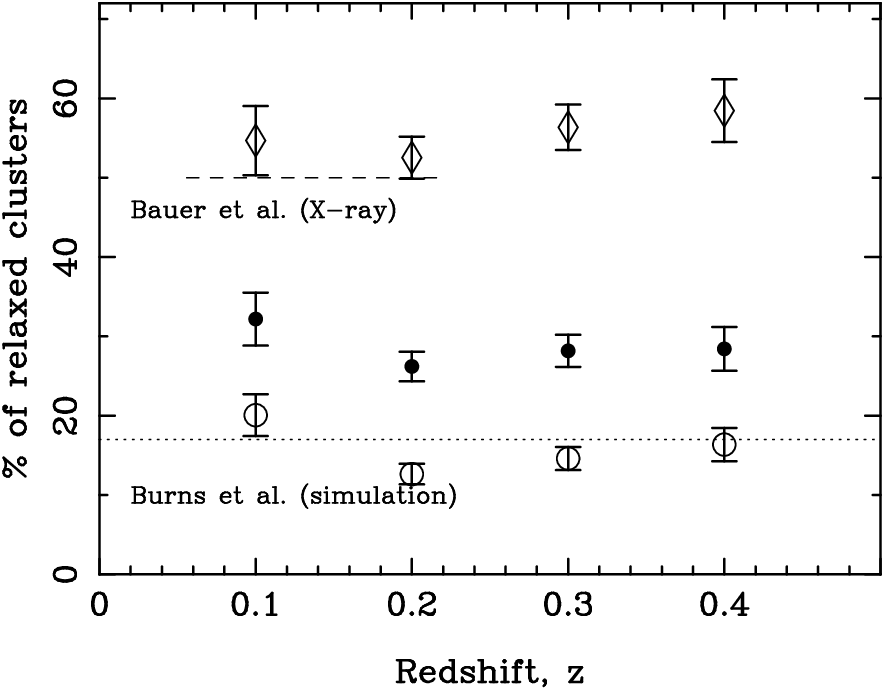}
\includegraphics[width = 7.5cm]{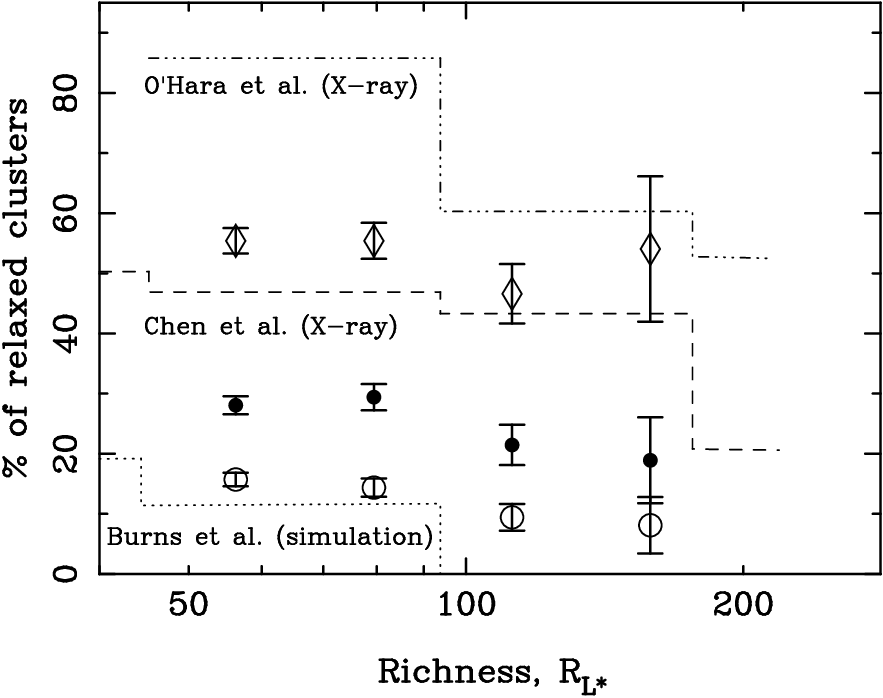}
\caption{Fraction of relaxed clusters does not vary with redshift
  (upper panel) and richness (lower panel). The black dots are the
  fractions of relaxed clusters defined by $\Gamma\ge0$, open circles
  for $\Gamma\ge0.2$ and diamonds for $\Gamma\ge-0.4$. In the upper
  panel, result from previous simulation by \citet{bhg+08} is shown by
  the dotted line. The result from X-ray data by \citet{bfs+05} is
  shown by the dashed line. In the lower panel, previous result from
  simulations by \citet{bhg+08} is shown by the dotted line, and
  result from the X-ray data in \citet{omb+06} by dash-dotted line and
  that from X-ray data in \citet{crb+07} by the dashed line.}
\label{rexevo}
\end{figure}

If the relaxed clusters are defined as those having $\Gamma\ge0$,
28.2\% (589) of 2092 clusters are relaxed. A few clusters with a
non-Gaussian velocity distribution experience ongoing merger exactly
along the line of sight \citep{evn+12,rlr13} and can not be recognized
by our two-dimensional method (see Appendix). This fraction of relaxed
clusters should be taken as an upper limit. The fraction of relaxed
clusters does not significantly vary with redshift and richness
(Fig.~\ref{rexevo}), which is also found if the criterion for
relaxed clusters is changed to be $\Gamma\ge0.2$ or $\Gamma\ge-0.4$.

This constant fraction is very consistent to the result for a complete
X-ray sample of 108 clusters in $0.15<z<0.7$, from which \citet{me12}
found that 27 X-ray-luminous clusters are merger and that the merger
fraction does change at $z<0.4$ but starts to increases with redshift
at $z\sim0.4$. As clusters with cool cores are mostly relaxed
clusters, \citet{bfs+05} showed that the fraction of X-ray cool-core
clusters do not vary with redshift at $z<0.2$. Our result of no
redshift dependence of relaxation parameters is consistent with the
conclusion of \citet{bfs+05} and also numerical simulations of
\citet{bhg+08}. The fractions of cool-core clusters were claimed to
vary with cluster mass from X-ray data \citep{omb+06,
  crb+07,bhg+08}. Using optical spectroscopic data, \citet{evn+12}
found that richer clusters tend to have more substructures. However,
our results show almost no obvious richness dependence for the
fraction of relaxed clusters within $R_{L\ast}\ge 50$ (see lower panel
of Fig.~\ref{rexevo}). This inconsistency needs to be investigated
in future.

Our optical cluster sample is approximately volume-limited complete,
while the X-ray clusters are usually flux limited or flux selected.
Clusters with cool cores are more likely to be detected and
selected in X-ray because they have high X-ray peaks in the central
region \citep{hmr+10}. As shown by the simulations of \citet{emp11},
the flux-limited X-ray cluster samples are significantly biased for
clusters with cool cores. This selection effect is more serious at
lower redshift and lower mass, which may explain the higher fraction
($\sim$50\%) of cool-core clusters for flux-limited X-ray sample
\citep[e.g.,][]{crb+07} than the fraction (28\%) of relaxed clusters 
in our optical sample. 

To verify the selection effect of X-ray sample, we cross-match the
2092 clusters with the X-ray selected clusters in the {\it ROSAT} all
sky survey \citep{bvh+00,bsg+04}, and get 159 matches. Among this
X-ray detected subsample, 74 clusters (46.5\%) have $\Gamma\ge0$. This
fraction is very close to the fraction of cool-core clusters
\citep{bfs+05,crb+07} but significantly larger than 28.2\% for our
sample in Table~\ref{tab2}.

\begin{figure}
\centering
\includegraphics[width = 7.3cm]{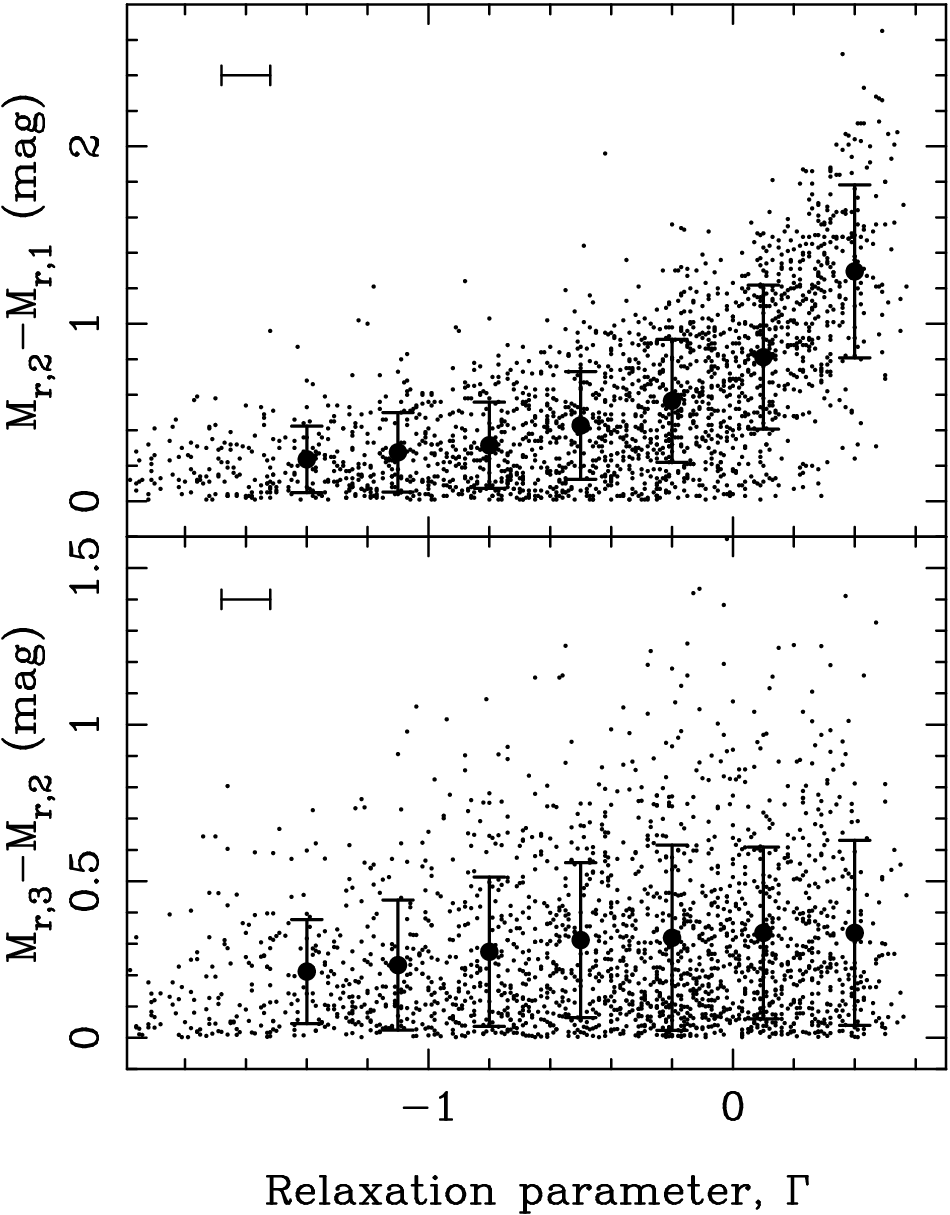}
\caption{Magnitude difference between the first BCG and second BCG
  (upper panel) and between the second BCG and third BCG (lower panel)
  against the relaxation parameter for 2092 rich
  clusters. The uncertainty of magnitude difference is less than 0.05
  mag, and the typical error-bar of $\Gamma$ is shown on the top
  left. The average and data scatter in seven bins are plotted to show the
  dependence.}
\label{mag12}
\end{figure}
\begin{figure}
\centering
\includegraphics[width = 7.3cm]{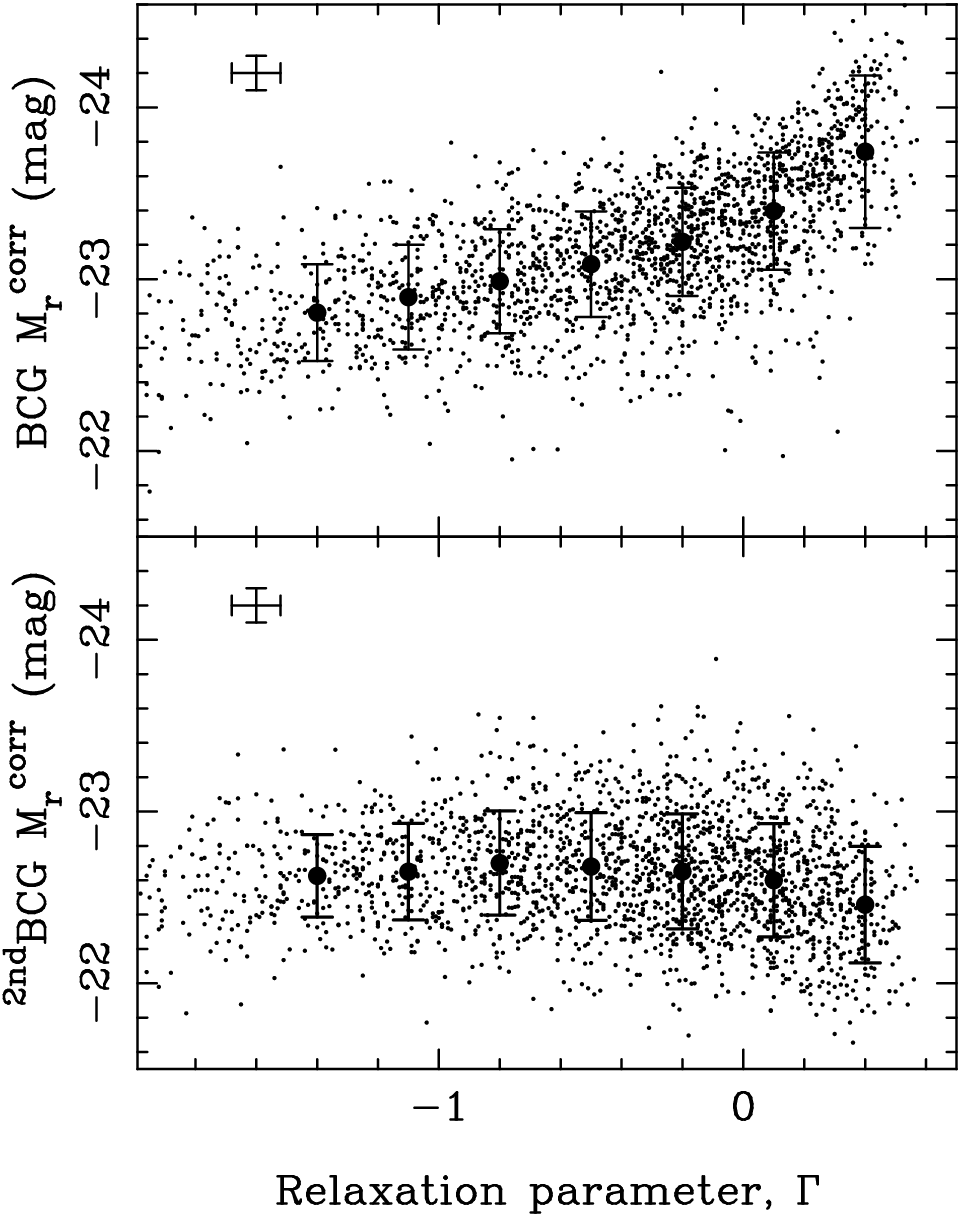}
\caption{Absolute magnitude of the first BCG (upper panel) and second
  BCG (lower panel) in the $r$ band, after the redshift evolution and
  richness dependence \citep{whl12} being diminished, against
  relaxation parameter. The cross bar on the top left indicates
  typical errors. The average and data scatter of the magnitudes in
  seven bins are plotted to show the dependence.}
\label{bcgmag}
\end{figure}
\subsection{Relaxation parameter and BCG dominance}

Very relaxed clusters in general have one very luminous BCG,
and unrelaxed clusters usually have more than one similarly brightest
member galaxies (see Fig.~\ref{exam3}). We now check if the BCG
dominance is related to cluster dynamical state quantified by the
relaxation parameter.

The BCG dominance is best shown by the difference of absolute
magnitudes of the first and second BCGs, e.g., $M_{r,2}-M_{r,1}$, 
in the $r$ band. As shown in the upper panel of
Fig.~\ref{mag12}, the magnitude difference obviously tends to be
larger for clusters with a larger relaxation parameter, which indicates
that the BCG dominance is closely related to dynamical states of
clusters \citep{rbp+07,skd+10}. However, the magnitude difference
between the second and third BCGs, $M_{r,3}-M_{r,2}$, does not
show significant dependence on relaxation parameter (the lower panel of
Fig.~\ref{mag12}).

We further check if more relaxed clusters have an absolutely more
luminous BCG, in addition to the relative BCG dominance. \citet{whl12}
noticed that BCG absolute magnitudes, after $k$-correction, evolves with
redshift and depends on richness. These effects have to be diminished
to show the dependence of absolute magnitude on dynamical state. We
correct these effects according to equations~7 and 9 in
\citet{whl12}, so that the corrected $r$-band absolute magnitude is
defined as
\begin{equation}
M_r^{\rm corr}=M_r+1.50\,z+1.10\,\log(R_{L\ast}/50).
\label{corrmag}
\end{equation}
As shown in the upper panel of Fig.~\ref{bcgmag}, the corrected BCG
absolute magnitude is related to the dynamical state. More relaxed
clusters host a more luminous first BCG \citep{skd+10}. The corrected
absolute magnitude of the second BCG is not obviously related to the
dynamical state (the lower panel of Fig.~\ref{bcgmag}).

We now conclude that BCG absolute magnitude and its relative dominance
are related to cluster dynamical state, in addition to the known
redshift evolution and dependence of cluster richness.

\begin{figure}
\centering
\includegraphics[width = 7.5cm]{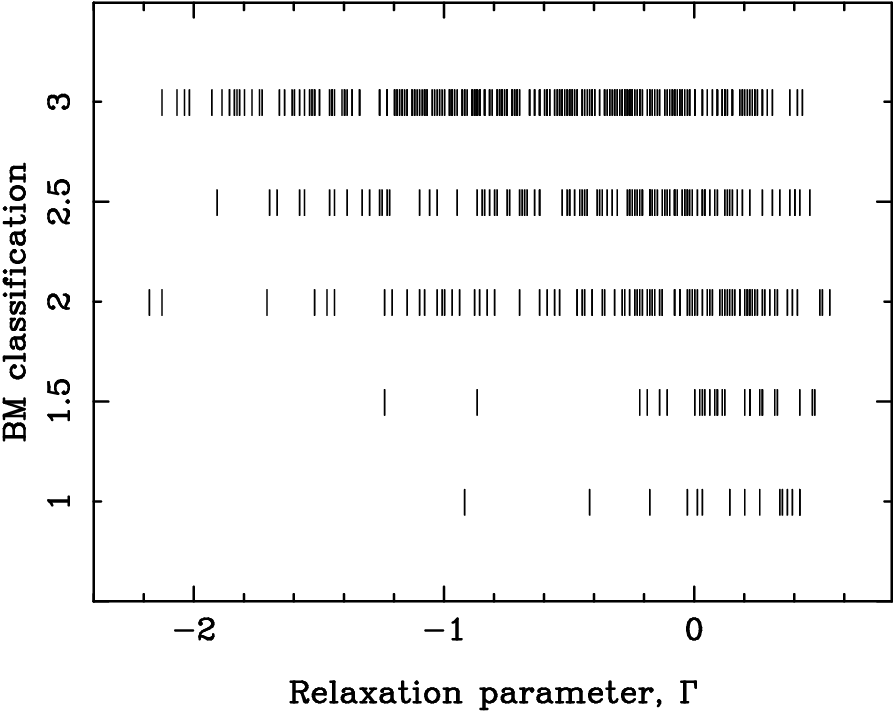}
\includegraphics[width = 7.5cm]{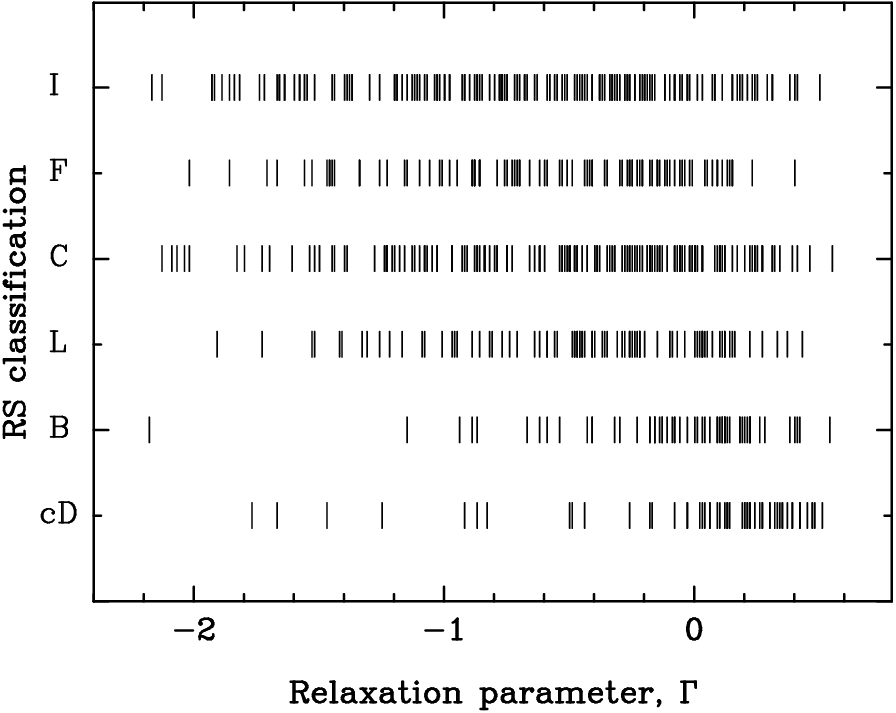}
\caption{Relaxation parameters for clusters with different 
BM cluster classification \citep[][upper panel]{bm70} and
with RS classification \citep[][lower panel]{rs71}.}
\label{bmrs}
\end{figure}
\subsection{Clusters classification and relaxation parameter}

Galaxy clusters have been classified according to galaxy distribution
in optical images \citep[see Table 1 of][for details]{bah77}. The
relaxation parameters we defined in this paper are derived from the
luminosity distribution of member galaxies. It is expected that
clusters of different types have different relaxation parameters.

Clusters are classified by \citet[][hereafter BM]{bm70} for five
types: I, I-II, II, II-III and III, based on the relative contrast of
the BCG. Type~I clusters contain a central cD galaxy, Type~II clusters
have a central galaxy between cD and Virgo-type giant ellipticals and
Type III clusters have no dominant galaxies. Type I-II and Type II-III
are the intermediate types between I and II and between II and III.
\citet{rs71} classified clusters into six types: cD, B, L, C, F and I,
based on the distribution of 10 brightest cluster member
galaxies. The cD-type clusters contain an outstandingly cD galaxy,
B-type clusters have two supergiant galaxies with a small separation,
L-type clusters have three or more supergiant galaxies among the top
10 brightest galaxies with comparable separation in a line, C-type
clusters have four or more brightest galaxies among the top 10 with
comparable separations in the core, F-type clusters have several
galaxies among the top 10 distributed in a flattened configuration,
and I-type clusters have the top 10 brightest galaxies distributed 
irregularly.

We cross-match clusters in Table~\ref{tab2} with the Abell clusters
\citep{aco89} which have BM classification. Within a separation of
$r_{200}$ and a redshift difference of 0.05, we get 509 matches.
Among them, 16, 28, 91, 102 and 272 clusters are of type I, I-II, II,
II-III and III, respectively. In the upper panel of Fig.~\ref{bmrs},
we show the $\Gamma$ distributions for the five BM types clusters.
About 75\% of type I and 79\% of type I-II clusters have
$\Gamma\ge0$, while this fraction decreases to 40\% for type II, 24\%
for type II-III and 14\% for type III clusters. The $\Gamma$
distributions also suggest that the type I and I-II clusters are more
relaxed, while type II to type III tend to be more unrelaxed,
consistent with the conclusion of \citet{bah77}.

There are 620 of 2092 clusters in Table~\ref{tab2} which have RS
classifications in \citet{sr87}. Among them, 58, 60, 85, 165, 91 and
161 clusters are classified as cD, B, L, C, F and I types,
respectively. The $\Gamma$ distributions for these types are plotted
in the lower panel of Fig.~\ref{bmrs}. The fractions of clusters
with $\Gamma\ge0$ for the cD and B type clusters (72\% and 52\%,
respectively) are significantly larger than those for the L, C, F and
I types (25\%, 22\%, 14\% and 14\%, respectively). The $\Gamma$
distributions suggest that cD- and B-type clusters tend to be more
relaxed, while F- and I-type clusters tend to be more unrelaxed.

\subsection{Dynamical state and radio halo}
\label{dynradio}

Radio halos are diffuse radio emission in clusters not
associated with any given member galaxies. \citet{bcd+09} found that
for clusters with detected radio halo, the radio power of halos is
closely related to the X-ray luminosity, by 
\begin{equation}
\log(P_{\rm 1.4 GHz}) -Y = A + b[\log(L_X)-X], 
\label{radx}
\end{equation}
where $P_{\rm 1.4 GHz}$ is the radio power at 1.4 GHz in unites of
W/Hz, $L_X$ is X-ray luminosity between 0.1 and 2.4~keV in unites of
erg s$^{-1}$, $Y=24.5$, $X=45$, $A=0.195\pm0.060$ and
$b=2.06\pm0.20$. However, low-frequency search for radio halos for
some X-ray luminous clusters failed to detect radio halos
\citep{vgd+08}. The non-detection suggests the bimodality for the
relation between X-ray luminosity and radio power \citep{bcd+09}. On
the other hand, radio halos are exclusively detected from merging
clusters \citep{buo01,ceg+10}, but there is no quantitative relation
between radio power and the degree of the cluster disturbance. The
information of cluster merger is expected to account for the scatter
and the bimodality of the $P_{\rm 1.4 GHz}$--$L_X$ relation. Here, we
quantitatively investigate the correlation between the deviation of
radio power from the $P_{\rm 1.4 GHz}$--$L_X$ relation with the
relaxation parameter.

We get the data for radio halo powers and X-ray luminosities of 15
clusters from \citet{fgg+12}: A665, A697, A746, A773, A781, A851,
A1351, A1689, A1758, A1914, A1995, A2034, A2219, A2255 and A2256.  All
these clusters are listed in our Table~\ref{tab1}. We get data for six
clusters with mini halos, A1835, A2029, A2142 and RX J1504.1$-$0248
from \citet{fgg+12} and A2390 and Z7160 from \citet{bcd+09}, because
the mini-halos also follow the same X-ray and radio relation. In
addition, we include data of 16 X-ray luminous clusters of $R_{L\ast}
\geq 50$ which have only the upper limits of radio halo powers because
of non-detection of radio halos. The data are obtained for 12
clusters: A611, A1423, A2537, A2631, A2697, MACS J1115.8+0129,
MACS J2228.5+2036, RX J0027.6+2616, RX J1532.9+3021, Z2701, Z5699 and
Z7215 from \citet{bcd+09} and for four clusters: A267, A1576, A2261
and RXC J0437.1+0043 from \citet{kvg+13}. All these clusters have been
included in Table~\ref{tab2}.

\begin{figure}
\centering
\includegraphics[width = 8.cm]{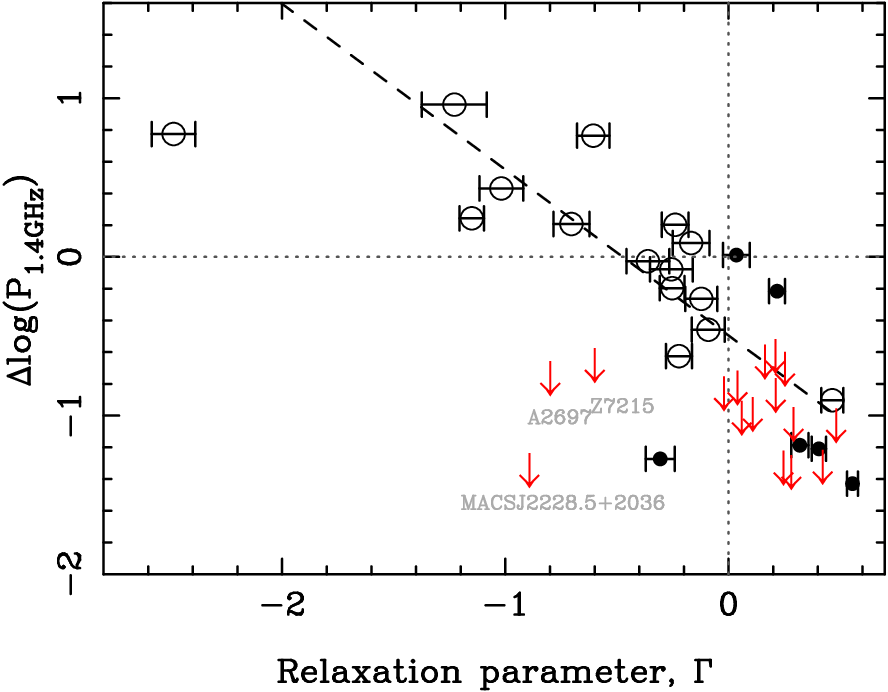}
\caption{Deviation of radio halo power from the X-ray and radio
  relation is very closely related to the relaxation parameter, $\Gamma$.
  Radio power and X-ray luminosity are taken from by \citet{fgg+12}
  and \citet{bcd+09} for radio halo clusters (open circles) and
  mini-halo clusters (black dots). The arrows are the upper limits of
  radio halo detection.  The dashed line indicates the best fit.
}
\label{pradio}
\end{figure}

With the relation between X-ray luminosity and radio halo power, as
shown in Equation~\ref{radx}, the radio halo powers of clusters can be
predicated from the observed $L_X$. The deviations of the observed
radio powers, $\Delta \log(P_{\rm 1.4 GHz})$, from the predictions are
then plotted against the relaxation parameter, $\Gamma$, in
Fig.~\ref{pradio}. We find a good correlation between the
deviations and relaxation parameter, given by
\begin{equation}
\Delta \log(P_{\rm 1.4 GHz})=(-0.49\pm0.11)-(1.05\pm0.19)\,\Gamma,
\end{equation}
for both the halos and mini-halos. Most X-ray luminous clusters with
non-detection of radio halos are relaxed clusters. Their upper limits of
radio powers are very close to the correlation line, except three
outliers (A2697 with $\Gamma=-0.80\pm0.08$, MACS J2228.5+2036 with
$\Gamma=-0.90\pm0.11$ and Z7215 with $\Gamma=-0.60\pm0.11$). Our
result suggests that dynamical states of clusters are the main
reason for the data scatter around the $P_{\rm 1.4GHz}$--$L_X$
relation in Fig. 8 of \citet{fgg+12}. This is the first time of
quantitative demonstration that radio halo is not only related to
X-ray luminosity but also to the dynamical state of clusters.

\begin{figure}
\centering
\includegraphics[width = 8.cm]{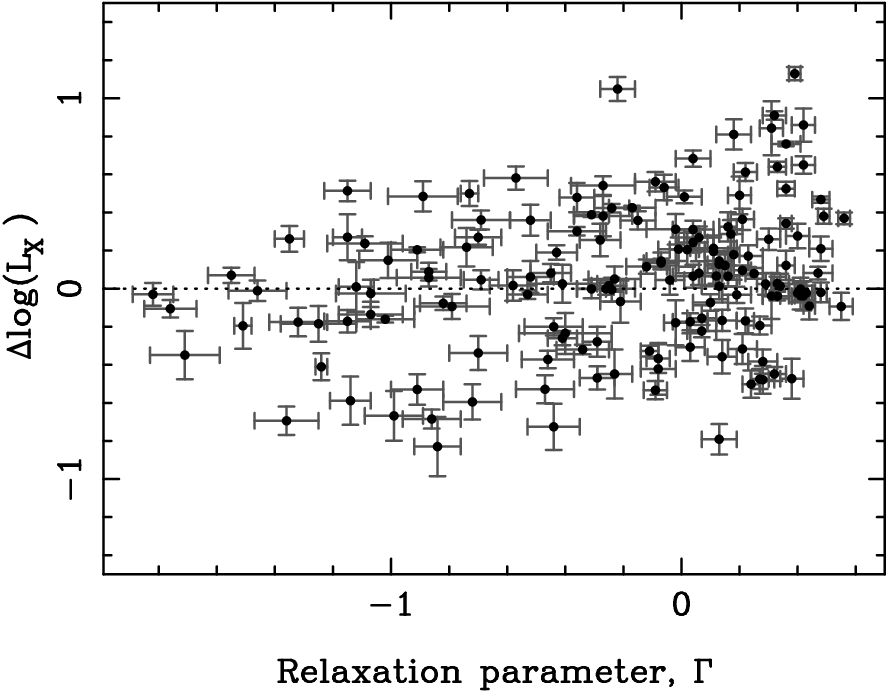}
\caption{For 159 X-ray clusters in our sample detected by the {\it ROSAT}
  all sky survey, the deviations of X-ray luminosity from the
  richness--$L_X$ relation given by \citet{whl12} are not correlated
  with the relaxation parameter, $\Gamma$.  The dotted line indicates
  $\Delta \log(L_X)=0$.}
\label{richlx}
\end{figure}

\subsection{Dynamical state and X-ray luminosity}

X-ray luminosities of clusters are tightly correlated with
cluster masses \citep{crb+07}. A good proxy of cluster masses is
cluster richness defined as $R_{L\ast}=L_{200}/L^{\ast}$ in
\citet{whl12} (also see Section~\ref{radrich}). The correlation
between cluster richness and X-ray luminosity given by \citet{whl12}
is
\begin{equation}
\log(L_X)-44.0=-2.49+1.59\log(R_{L\ast}).
\label{xr}
\end{equation}
The data have fairly scatter around this relation \citep[see fig. 17
  of ][]{whl12}, but the reason is not clear. \citet{pbb+07b} showed
that clusters in ongoing-merging process have a low X-ray luminosity,
which implies that the dynamical states of clusters could influence
the X-ray luminosity. Their study suggests that cluster dynamical
state may account for the scatter of the $L_X$--$R_{L\ast}$
relation. Here, we check if the deviation of X-ray luminosity from the
$L_X$--$R_{L\ast}$ relation is related to the relaxation parameter.

In our Table~\ref{tab2}, 159 clusters have been detected in the {\it
  ROSAT} X-ray all sky survey \citep{bvh+00,bsg+04}. Now, we assume
that cluster masses, and hence cluster richnesses, are fundamentally
related to cluster X-ray luminosity. Therefore, we can predict an
X-ray luminosity from the richness by using Equation~(\ref{xr}), and
then get the offset between the predicted and observed X-ray
luminosity, $\Delta \log(L_X)$. As shown in Fig.~\ref{richlx}, we
find very week correlation between $\Delta \log(L_X)$ and $\Gamma$,
which suggests that the global X-ray luminosity is insensitive to
cluster dynamical state.

\section{Conclusions}

We presented a robust method to diagnose substructures and dynamical
states of galaxy clusters by using the optical photometric data of
member galaxy distribution.  The distribution of member galaxies is
smoothed by using a Gaussian kernel and a weight of their
luminosities. The asymmetry factor, $\alpha$, the ridge flatness,
$\beta$, and the normalized deviation, $\delta$, are then calculated
from the smoothed map, based on which a relaxation parameter,
$\Gamma$, is defined to quantify dynamical states of clusters. The
smooth-scale and parameter combination are then optimized by using a
test sample of 98 clusters with known dynamical states previously
classified as relaxed and unrelaxed based on X-ray, optical and radio
data. The newly defined relaxation parameter, $\Gamma$, can be used to
distinguish the known relaxed and unrelaxed clusters with a success
rate of 94\%, only a few exception of mergers along the line of sight.

We calculated relaxation parameters for 2092 clusters in \citet{whl12}
with a richness of $R_{L\ast}\ge50$ identified from SDSS. We found
that the relaxation parameters are continuously distributed in a range
of $-2\lesssim\Gamma<0.6$. Only 28\% of 2092 rich clusters are
classified as relaxed clusters with $\Gamma\ge0$. This fraction is
smaller than that of the matched X-ray subsample detected by the
{\it ROSAT}, which confirms that the flux-selected X-ray cluster sample
usually has a selection bias in dynamical state. The fraction of
relaxed clusters does not vary significantly with redshift at
$z\le0.42$ and with richness in the range of $R_{L\ast}\ge50$ (i.e.,
$M_{200}\ge3.15\times10^{14}~M_{\odot}$). Our results imply that a
large fraction of clusters are still continuously growing even for
massive ones. We found that the relaxation parameter strongly
correlates with the absolute magnitude of BCGs and with the magnitude
difference between the first and second BCGs, which indicates that BCG
growth is related to dynamical state of its host cluster.  For the
first time, we quantitatively showed that the emission power of radio
halo not only depends on the X-ray luminosity but also the dynamical
state of a cluster.

\section*{Acknowledgements}

The authors are supported by the National Natural Science Foundation
of China (10833003 and 11103032) and the Young Researcher Grant of
National Astronomical Observatories, Chinese Academy of Sciences.
Funding for SDSS-III has been provided by the Alfred P. Sloan
Foundation, the Participating Institutions, the National Science
Foundation and the US Department of Energy.  The SDSS-III web site is
http://www.sdss3.org/.
SDSS-III is managed by the Astrophysical Research Consortium for the
Participating Institutions of the SDSS-III Collaboration including the
University of Arizona, the Brazilian Participation Group, Brookhaven
National Laboratory, University of Cambridge, University of Florida,
the French Participation Group, the German Participation Group, the
Instituto de Astrofisica de Canarias, the Michigan State/Notre
Dame/JINA Participation Group, Johns Hopkins University, Lawrence
Berkeley National Laboratory, Max Planck Institute for Astrophysics,
NewMexico State University, New York University, Ohio StateUniversity,
Pennsylvania State University,University of Portsmouth, Princeton
University, the Spanish Participation Group, University of Tokyo,
University of Utah,Vanderbilt University, University of Virginia,
University of Washington, and Yale University.

\begin{appendix}
\section{Five clusters with known dynamical states but unusual relaxation parameters}
\label{note}

Most clusters with known dynamical states can be separated by
the plane in Fig.~\ref{abc} for their relaxed states or unrelaxed
states. However, five clusters have unusual relaxation parameters,
CL0024.0+1652, A267, A370, A1689, A1991. Here, we investigate them for
details.

CL0024.0+1652: we get a relaxation parameter of
$\Gamma=0.49\pm0.04$, which means `very relaxed'. It has a regular
morphology in X-ray image, and does not show a cool core. There is
no dominant central galaxy at the center
\citep{bsm+00,zbm+05}. However, the optical spectroscopic data of
member galaxies show a non-Gaussian redshift distribution, which
indicates that CL0024.0+1652 is an ongoing merger along the line of
sight \citep{cmk+02}. It is therefore not surprising that the dynamical
state of CL0024.0+1652 can not be figured out from optical luminosity
distribution of member galaxies.

A267: it has a dominant central galaxy with magnitude difference
between the first and second BCGs, $M_{r,2}-M_{r,1}=2.14$. We get a
very regular smoothed optical map and a high relaxation parameter of
$\Gamma=0.48\pm0.02$. However, the X-ray and lensing measurements show
that A267 has a large offset between the mass centered on BCG and
X-ray peak \citep{sks+05}, which indicates unrelaxed dynamical state
of this cluster. No cool core is found \citep{bfs+05}.

\begin{figure}
\centering
\includegraphics[width = 7.5cm]{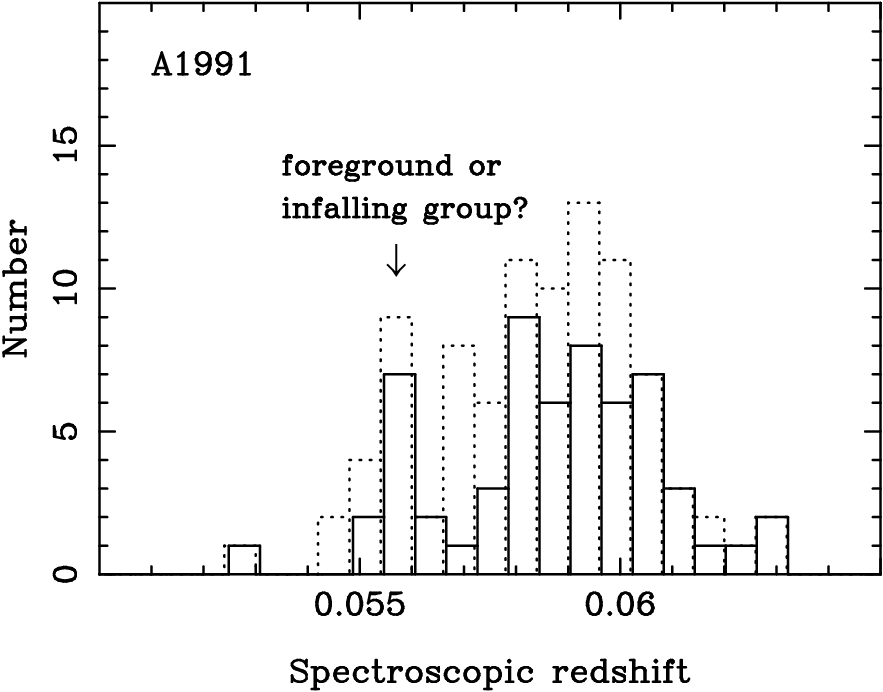}
\caption{Redshift distribution of member galaxies within $r_{200}$ (dotted
  line) and $r_{500}$ (solid line) for A1991. The data are extracted
  from the SDSS-III.}
\label{A1991}
\end{figure}

A370: using optical data, we get a relaxation parameter of
$\Gamma=0.07\pm0.05$. \citet{omf98} suggested that A370 consists of
two subclusters in the light-of-sight direction. Two mass clumps are
figured out by strong lensing centered on two BCGs with a projected
separation of 200 kpc and a velocity difference of about 1000
km~s$^{-1}$ \citep{kmf+93}. X-ray image shows two peaks centered on two
BCGs \citep{sqf+10}, and there is no cool core in X-ray \citep{me07}.

A1689: we get a high relaxation parameter of
$\Gamma=0.47\pm0.05$, and it shows a regular morphology in X-ray image
and looks like a relaxed cluster \citep{xw02}. However, the
temperature profiles show evidence for merger along the line of sight
\citep{am04}. In X-ray, A1689 has a cool core
\citep{all00,crb+07}. In radio, a small radio halo has been detected
in the central region of this cluster \citep{vgm+11}.

A1991: using optical photometric data, we get a relaxation
parameter of $\Gamma=-0.41\pm0.06$ for this cluster. It has a regular
morphology and cool core in X-ray \citep{vmm+05, wok+10}. However,
spectroscopic data of member galaxies show the non-Gaussian velocity
distribution (see Fig.~\ref{A1991}). The feature may suggest that a
group is infalling into the cluster \citep{smn+04} or the presence of
a possible foreground structure.

\end{appendix}

\label{lastpage}
\end{document}